\documentclass[manuscript,screen]{acmart}

\usepackage{xcolor}
\usepackage{algorithmic}
\usepackage{graphicx}
\usepackage{subcaption}
\usepackage{textcomp}
\usepackage{comment}
\usepackage{hyperref}
\usepackage{bm}
\usepackage{tcolorbox}{\small}
\usepackage{threeparttable}

\usepackage{lipsum}
\usepackage{multirow}
\usepackage[justification=centering]{caption}
\usepackage{caption}
\usepackage{url}
\usepackage{listings}
\usepackage{colortbl}
\usepackage{enumitem}
\usepackage{soul}
\usepackage{listings}
\usepackage{makecell}
\usepackage{float}
\usepackage{mdframed}
\usepackage{listings}
\hypersetup{hidelinks}

\begin{document}

\title{Vision to Specification: Automating the Transition from Conceptual Features to Functional Requirements}

\author{Xiaoli Lian}
\email{lianxiaoli@buaa.edu.cn}
\affiliation{%
  \institution{SKLCCSE, School of Computer Science and Engineering, Beihang University}
  \city{Beijing}
  \country{China}
}

\author{Jiajun Wu}
\email{1074275896@qq.com}
\affiliation{%
  \institution{School of Computer Science and Engineering, Beihang University}
  \city{Beijing}
  \country{China}}

\author{Xiaoyun Gao }
\email{fishdog@buaa.edu.cn}
\affiliation{%
 \institution{School of Computer Science and Engineering, Beihang University}
 \city{Beijing}
 \country{China}
}

\author{Shuaisong Wang}
\email{littletree@buaa.edu.cn}
\affiliation{%
  \institution{School of Computer Science and Engineering, Beihang University}
  \city{Beijing}
  \country{China}}
  
\author{Li Zhang}
\email{lily@buaa.edu.cn}
\affiliation{%
  \institution{SKLCCSE, School of Computer Science and Engineering, Beihang University}
 \city{Beijing}
 \country{China}
 \authornote{Xiaoyun Gao and Li Zhang are the corresponding authors.}
 }

\renewcommand{\shortauthors}{Xiaoli Lian et al.}

\begin{abstract}
The translation of high-level abstract features into clear, and testable functional requirements (FRs) is a crucial step in software development, bridging the gap between user needs and technical specifications. In engineering practice, significant expert effort is needed for this translation. Our approach, \textit{EasyFR}, streamlines the process by recommending Semantic Role Labeling (SRL) sequences for the given abstract features to guide Pre-trained Language Models (PLMs) in producing cohesive FR statements. By analyzing ten diverse datasets, we induce two variable SRL templates, each including two configurable parts. For concrete features, our proposed \emph{Key2Temp} model can construct the appropriate variant of the SRL template by identifying a variable SRL template and placing the feature tokens in the appropriate slots. In this way, our approach reframes the process of requirement generation into a structured slot-filling activity.
Experimental validation on four open datasets demonstrates that \textit{EasyFR} outperforms three advanced Natural language generation (NLG) approaches, including GPT-4, particularly when existing FRs are available for training. The positive influence of our SRL template variant recommendations is further confirmed through an ablation study. We believe that our results indicate a notable step forward in the realm of automated requirements synthesis, holding potential to improve the process of requirements specification in future software projects.
\end{abstract}





\maketitle

\section{Introduction}
\label{sec:intro}

In the quest to deliver software solutions that effectively meet user needs and business objectives, specifying clean, clear, and testable functional requirements (FRs) is paramount \cite{Terzakis2013TheIO, REManaging, REMicrosoft}. The genesis of these concrete FRs typically starts with a vision, i.e., a conceptual understanding of desired features at a high level of abstraction \cite{10.5555/326459}. These envisioned features articulate capabilities \cite{DABROWSKI2023102181}, user-visible characteristics \cite{9793551}, or quality aspects \cite{DBLP:journals/ese/JhaM19} (e.g., lightweight) from different dimensions. In this work, we refer to a feature as a distinctive functionality or capability. Features are often framed in broad terms with high-level descriptions that lack detailed specifics, such as short sentences like ``RBC controls train movements'' or single phrases like ``sending videos'' \cite{featureMei}. Obviously, navigating from this elevated vision of features to finely-tuned FR specifications entails intricately translating general notions into actionable requirements amenable to systematic implementation and rigorous testing by development teams.

Transitioning from abstract visions to specific functionalities presents challenges that can confront even the most experienced requirements engineers. Initially, each feature must be broken down into its constituent functionalities, which demands an extensive understandingding of the domain space and the technical landscape. Engineers must draw upon pertinent domain knowledge and the unique context of the project to interpret the features correctly and demarcate the requirements boundaries \cite{DBLP:conf/re/LianRCZFS16, 10.5555/326459, 10.5555/113586}. Furthermore, they need to incorporate the necessary detail to enhance requirement clarity. In addition, the articulation of specifications requires adept writing capabilities, the production of documents that adhere to the syntactic norms of FRs while containing sufficient information for subsequent development and testing phases, all expressed with lucid and exact language. Given these multifaceted demands, the specification of FRs heavily relies on the domain knowledge of involved experts, which ultimately determines the quality of the FRs. Furthermore, because this process requires iterative human thinking, crafting, and modification, it is time-consuming. Therefore, an automated approach is necessary.

While prior work has extensively explored the automatic conversion of structured and semi-structured models into natural language (NL) requirements \cite{DBLP:journals/re/MaidenMJG05, DBLP:conf/coopis/YuBDM95, DBLP:conf/sigsoft/LetierL02, DBLP:journals/re/LandtsheerLL04, van2004goal, DBLP:journals/tse/LamsweerdeW98, DBLP:journals/re/MezianeAA08, DBLP:conf/re/Berenbach03a, 9793770},  our literature review did not identify studies specifically addressing automated translation of abstract features to concrete, implementable FRs. Our work bridges this gap by proposing the first approach to automatically synthesize concrete FRs from high-level features using pre-trained language models (PLMs).

Our focus lies on FRs that define the behavior of a system. As observed, ``most products and applications, conceived to perform useful tasks, are rich with software functional requirements'' \cite{ManagingRE}. These FRs form the foundation for products, and they also serve as the groundwork for non-functional requirements. Our approach strives to reduce the manual effort involved in preparing and drafting specifications. PLMs, particularly those designed for general-purpose use, are known to encapsulate a vast array of domain knowledge \cite{10.5555/3495724.3495883}. Therefore, the primary challenge for the PLM-driven methodologies lies in automatically retrieving relevant domain knowledge, integrating it with the provided features, and crafting coherent natural language statements that conform to the clarity and structure typically expected in FRs specifications.

Our solution introduces Semantic Role Labeling (SRL) templates as a structured bridge between abstract features and concrete requirements. These templates consist of configurable sequences of frequently occurring SRL roles found in FRs specifications. Crucially, we recommend a distinct SRL template variant for each feature, a customized configuration of the template tailored to the feature's specific semantic roles, enabling PLMs to generate more focused and relevant requirement statements.

Our design is motivated by two principal reasons. First, shaping requirements for a specific feature is a constrained natural language generation (NLG) task that must navigate both the context provided by the feature and the conventional norms of requirements expression. The SRL template we have formulated offers a structured framework that captures the essential semantic elements necessary for crafting a FR. It incorporates those elements already present within the given features and facilitates the generation of new ones. Consequently, this aids in the requirement synthesis process by allowing for the systematic filling of semantic placeholders in the template. In addition, our SRL templates encompass a variety of semantic roles that articulate the different abstract functions fulfilled by predicate arguments within event contexts. This structure complements the syntactical constructs of FRs, as demonstrated by the widely adopted ISO/IEC/IEEE 29148:2018 standard \cite{8559686}, henceforth referred to as ISO 29148.

\begin{figure}[htbp]
	\centering
	\includegraphics[trim = {0cm 0.5cm 1.5cm 1.5cm}, clip, width=\textwidth]{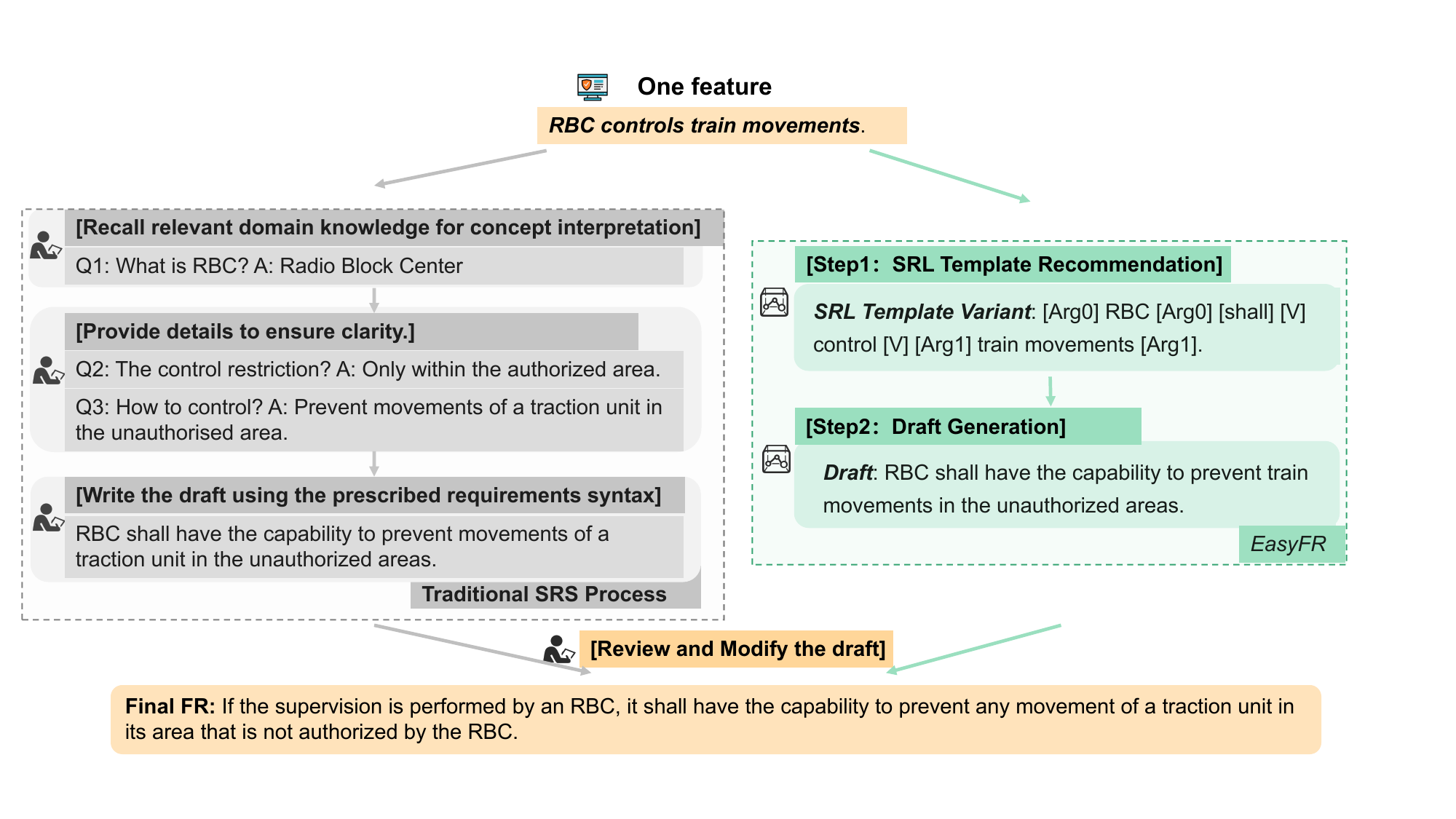}
	\caption{An illustrative example showing our target and design.}
	\label{fig:MotiExample}
\end{figure}

We demonstrate our methodology and its application via an illustrative example in Fig. \ref{fig:MotiExample} and juxtapose it against the traditional process of requirement formulation for comparative analysis. Note that the traditional process is depicted according to the typical procedure for specifying software requirements. It may be incomplete for some complex scenarios, such as those involving overly abstract features (e.g., single feature token) or complex requirements that demand multi-sentence descriptions.

In detail, we introduce \textit{EasyFR}, a novel approach designed specifically for the generation of FRs drafts from defined features. In the initial phase, we manually discerned eight SRL roles that are predominantly utilized and extracted two variable SRL templates from a corpus of 768 requirements across ten publicly accessible datasets. Each identified SRL template consists of several fixed elements and incorporates two parts that are subject to variation. Subsequently, we developed a model, \emph{Key2Temp}, which facilitates the recommendation of a fitting SRL template by selecting a variable template and instantiating its variable components. Finally, we fine-tuned the generative language model ChatGLM-6B\footnote{\url{https://github.com/THUDM/ChatGLM-6B}}\cite{team2023chatglm} to automate the drafting of FR documents. To demonstrate our approach more clearly, we provide an example in the right zone of Fig. \ref{fig:MotiExample}. By recasting the constrained task of NLG for FRs into an SRL slot-filling exercise, our template-based method greatly simplifies the generation of preliminary FR drafts.

Our experiments across four open datasets reveal that \textit{EasyFR} delivers impressive results, even surpassing the performance of three state-of-the-art NLG methods, GPT-4 included. To reflect real-world use cases where a requirements assistant might need to generate requirements either from scratch or based on existing ones from current or similar projects, we assess \textit{EasyFR} in two distinct scenarios. Our findings indicate that \textit{EasyFR} performs on par with or even better than the sophisticated GPT-4 model when there are no pre-existing requirements to learn from. Furthermore, \textit{EasyFR} excels beyond GPT-4's capabilities when it has access to pre-existing requirements for training. We also highlight the beneficial impact of SRL template recommendations through an ablation study, reinforcing the value of this feature in our approach.

The main contributions of this work include:
\begin{itemize}[leftmargin = 0.5cm]
\item To the best of our knowledge, \textit{EasyFR} is the first publicly available approach capable of automatically transitioning from abstract features to concrete FR statements.
\item Proposing two variable SRL templates and eight common SRL roles from 768 requirements, along with one trained model \textit{Key2Temp}, capable of recommending configured SRL templates variant for given features or tokens for FR.
\item Evaluating \textit{EasyFR} across four datasets in two scenarios - with / without preexisting requirements. This demonstrates the efficacy of our \textit{EasyFR} and the positive impact of the recommendations of the SRL template.
\end{itemize}

The rest of this paper is organized as follows. Section \ref{sec:background} provides the background information on FR syntax in ISO/IEC/IEEE 29148:2018, SRL, and two related PLMs. Section \ref{sec:approach} describes our proposed approach, \emph{EasyFR}. Section \ref{sec:eva} presents the experiments and results used to evaluate our approach.
Section \ref{sec:discussion} discusses the threats to validity and limitations of our approach.
Section \ref{sec:relatedWork} reviews related work in the field. Finally, Section \ref{sec:conclusion} concludes the paper and Section \ref{sec:dataAvailability} addresses data availability.
\section{Background}
\label{sec:background}

\subsection{Functional Requirements Syntax in ISO/IEC/IEEE 29148:2018}
\label{subsec:functionalReqSyntax}

ISO/IEC/IEEE 29148:2018 is a fundamental international standard for requirements engineering \cite{8559686}. This standard delineates the structure of FRs into five distinct components: [condition], [subject], [action], [object], and [constraint of action]. For enhanced clarity, we present an exemplary requirement from this standard: ``\emph{Upon receiving signal x{\footnotesize[Condition]}, the system{\footnotesize[Subject]} shall set{\footnotesize[Action]} the `signal x received' bit{\footnotesize[Object]} within 2 seconds{\footnotesize[Constraint of Action]}}''. This requirement is structured in a manner that mirrors the construction of a typical English sentence. Typically, the condition corresponds with the conditional adverbial clause, the subject and object within the requirement's syntax align with their counterparts in natural language sentences, the action relates to the verb or predicate, and the constraint functions akin to the sentence complement. Such parallelism with everyday language makes the standard's approach readily applicable in practical scenarios.
\begin{itemize}[leftmargin = 0.5cm]
    \item \emph{Condition}: The condition under which the requirement applies or is activated. Conditions can encompass environmental factors, system states, user actions, time constraints, and more. 
    \item \emph{Subject}: The subject generally denotes the system or a segment of the system (such as a module, component, or function) that carries out the action.
    \item \emph{Action}: This illustrates the action that the subject will execute. It could be operations like display, compute, send, receive, and so forth.
    \item \emph{Object}: The object typically represents the entity upon which the action is being performed.
    \item \emph{Constraint of Action}: These provide additional details or limitations concerning how the action should be executed or what results should be accomplished.
\end{itemize}



\subsection{Semantic Role Labeling (SRL)}
\label{subsec:SRL}

SRL\cite{marcheggiani-etal-2017-semantic} aims at understanding the predicate-argument structure of a sentence. It involves the identification of verbs (predicates) along with their associated entities (arguments) and the semantic relationships between them. Its principle is similar to that of FRs. According to Aurum and Wohlin \cite{REManaging}, essentially, FRs tell \emph{what the system will do}. The corner part is the \emph{actions} that have to take place in the software \cite{8559686} and the related arguments (e.g., the executor and restriction of action). This is the first reason that we choose SRL as the intermediate template.

The second reason is that SRL has rich, fine-grained semantic roles. Taking the commonly used Proposition Bank (PropBank) \cite{pradhan-etal-2022-propbank, palmer-etal-2005-proposition} as one example, there are three categories of semantic tags, including prediate (REL), core arguments (ArgN) and semantic adjuncts (ArgM). Predicate represents the core action. ArgN tags (e.g., Arg0, Arg1, Arg2) are typically used to facilitate computer understanding of who did what to whom in a sentence, which is crucial for many NLP tasks. ArgM tags (e.g., ArgM-LOC, ArgM-TMP, ArgM-MNR ...) are used for adjuncts or modifiers—optional constituents of a sentence that add extra information about the action. Given the similar principles to FRs, SRL's detailed semantic roles offer strong potential to further guide the generation of FR specifications.

We select AllenNLP\footnote{\url{https://github.com/allenai/allennlp}} to annotate the PropBank SRL on FR statements. We choose this tool for three reasons. Firstly, it offers a platform that combines the latest NLP techniques (e.g., Hugging Face’s Transformers\footnote{\url{https://github.com/huggingface/transformers}}), ensuring state-of-the-art performance. Secondly, it provides user-friendly tooling that simplifies the process of model training, evaluation, and deployment, making it accessible even for those who may not be NLP experts. Thirdly, the tool has strong community support, which includes extensive documentation, tutorials, and forums where users can seek help and share insights.


\subsection{Pre-trained Language Models (PLMs)}

Two PLMs are involved in our study, that is, Bidirectional Encoder Representations from Transformers (BERT) and ChatGLM-6B.

\subsubsection{Bidirectional Encoder Representations from Transformers (BERT)}

BERT, introduced by Google in 2018 \cite{devlin2019bert}, employs a purely encoder-based
Transformer architecture to capture bidirectional contextual information from large-scale unlabeled text. Unlike decoder-based models
(e.g., GPT \cite{radford2018gpt}), which process text from left to right, BERT ingests the entire sequence of words simultaneously,
capturing rich bidirectional contextual information during both pre-training and fine-tuning.
Pre-trained on massive unlabeled corpora through self-supervised tasks---namely Masked Language Modeling (MLM),
which masks random tokens and predicts them from surrounding context, and Next Sentence Prediction (NSP),
which determines if two segments of text are consecutive---BERT learns contextual dependencies across entire
sentences \cite{devlin2019bert}. This bidirectionality allows BERT to discern subtle, complex patterns within sequences, making it
particularly adept for tasks requiring nuanced comprehension \cite{hashemi2020realtime,  kim2022malware}.

Once pre-trained, BERT can be efficiently fine-tuned for downstream tasks by optimizing all parameters with a simple task-specific layer. This adaptability enables strong performance across diverse NLP applications, including classification tasks and syntactical relationship modeling—such as determining semantic roles for feature tokens.
 

\subsubsection{ChatGLM-6B}
\label{subsubsec:chatglm}

ChatGLM-6B, developed by Tsinghua University, is a bilingual (Chinese and English) conversational language model that
builds on the General Language Model (GLM) architecture \cite{team2023chatglm}. In contrast to the encoder-only design of BERT, ChatGLM-6B
adopts a decoder-only Transformer architecture. Its pre-training corpus consists of diverse and multilingual data drawn from a variety of sources
(e.g., webpages, books, research papers), augmented by high-quality filtering and deduplication procedures to enhance
data diversity and overall model robustness \cite{du2022glm}. Tokenization is handled through a merged vocabulary of Chinese and
multilingual byte pair encodings, allowing ChatGLM-6B to support a broad range of lexical items across languages.

Following pre-training, ChatGLM-6B undergoes alignment and post-training steps to improve its adherence to human
preferences and task-specific instructions. Leveraging Supervised Fine-Tuning (SFT) and Reinforcement Learning from
Human Feedback (RLHF), ChatGLM-6B learns to address a range of nuanced tasks and multi-turn dialogues more reliably.
Additionally, techniques such as Rotary Positional Encoding (RoPE) and strategies like Group Query Attention (GQA)
are utilized to manage large context lengths and facilitate efficient inference. To further reduce computational overhead,
ChatGLM-6B benefits from partial-parameter fine-tuning methods like P-Tuning v2. 

P-Tuning v2 has been developed to facilitate downstream developers in customizing the model for their own application scenarios, yielding the performance of fine-tuning while having only 0.1\%-3\% tuned parameters \cite{liu-etal-2022-p}. Thus, it can substantially reduce memory cost during model training. As an optimized implementation of traditional Deep Prompt Tuning \cite{li-liang-2021-prefix}, P-Tuning v2 serves as a lightweight alternative to fine-tuning for NLG tasks. It keeps language model parameters frozen and optimizes a sequence of continuous task-specific vectors. Unlike Deep Prompt Tuning, which applies continuous prompts at the input layer, P-Tuning v2 applies them at every layer of the pre-trained model, enhancing its adaptability for NLG tasks.

These technical capabilities establish a robust foundation for our approach, enabling the generation of high-quality FRs from feature tokens and recommended SRL templates (Section \ref{subsec:draftGeneration}).


\section{Approach}
\label{sec:approach}

\begin{figure*}[htbp]
	\centering
	\includegraphics[trim={0cm 2.5cm 0.5cm 1.5cm}, clip, width=\textwidth]{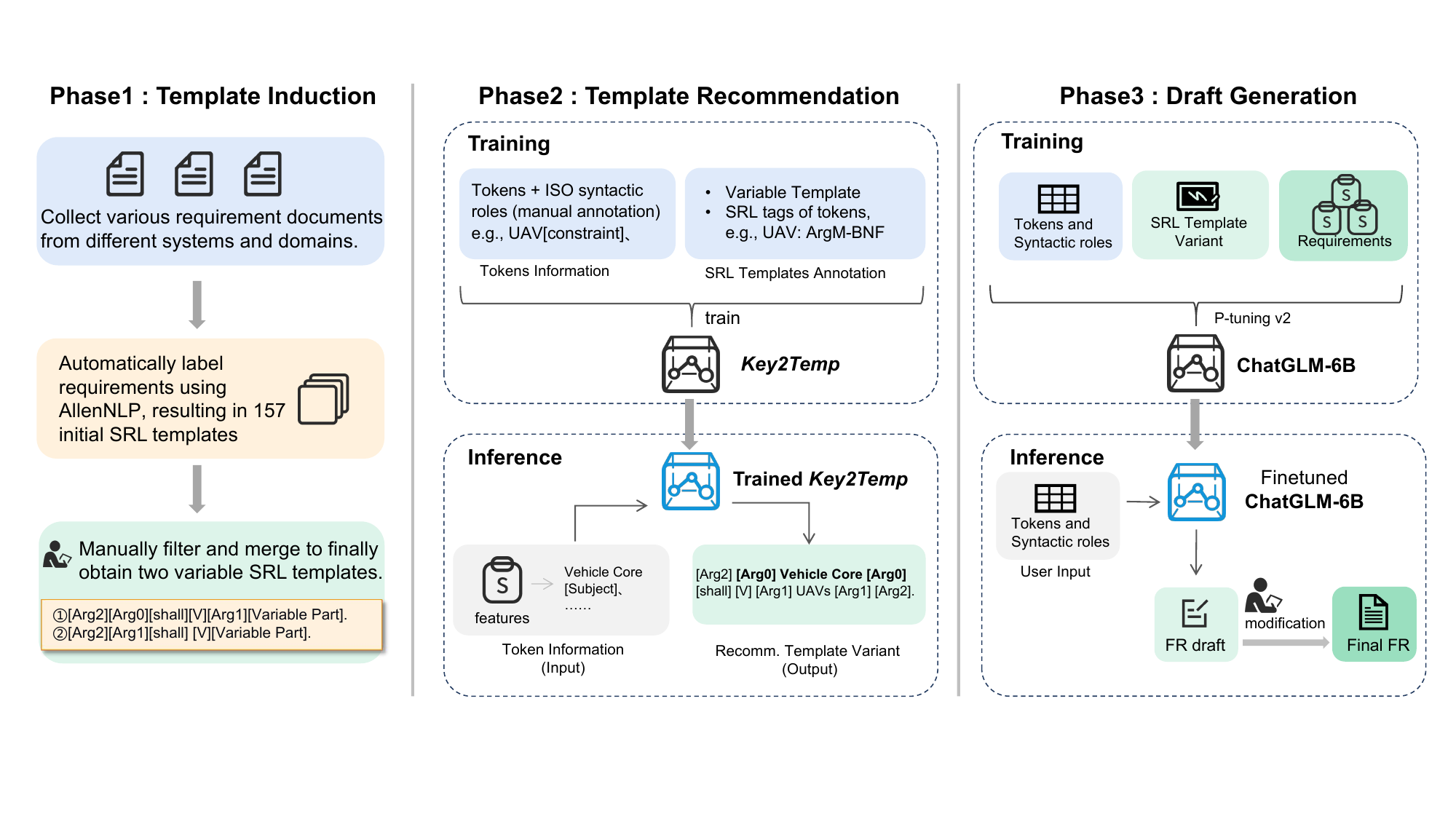}
	\caption{Overview procedure of our approach \textit{EasyFR}.}
	\label{fig:overview}
\end{figure*}

\textit{EasyFR} is structured into three sequential stages: template induction, template recommendation, and draft generation, as depicted in Fig. \ref{fig:overview}. Initially, we engage in template induction by manually identifying commonly used SRL templates from a compilation of 10 requirements datasets. In the subsequent stage, we introduce \textit{Key2Temp}, a model trained to automatically suggest the most fitting SRL template based on the analyzed feature. To facilitate this recommendation process, features are dissected into individual semantic units called tokens, along with their syntactic roles as prescribed by ISO 29148. These tokens can be single words, phrases, or clauses. In the final phase, we utilize P-Tuning v2 \cite{liu-etal-2022-p} to fine-tune the ChatGLM-6B model, the backbone of our \textit{EasyFR}. Incorporating the SRL templates proposed by \textit{Key2Temp}, \textit{EasyFR} is then able to efficiently generate FRs towards the given features. 

\subsection{Phase I: Template Induction}
\label{subsec:tempalteReduction}
The aim of this phase is to extract frequently occurring semantic templates from publicly available requirement datasets. Here, a `template' refers to a consistent pattern in the sequence of SRL tags that recur across various FRs statements.

We selected 10 distinct sets of open requirements, including GANNT \cite{8049173}, WARC \cite{8049173}, and eight in HIPAA (including Care2x, ClearHealth, iTrust, PatientOS, PracticeOne, WorldVista, Consultations, and Soren) \cite{HIPPA} which collectively encompass 975 individual requirements. Each set was constructed by a different group. Although the requirement documents under consideration were sourced from only three domains (i.e., project management, web archive, and healthcare), they present variation conducive to identifying common SRL templates. Firstly, various organizations adhere to distinct syntactical conventions \cite{8804292}, with documentation across different entities often reflecting unique stylistic characteristics. To capture this diversity, we have incorporated documents from 10 different organizations. This approach ensures that our analysis represents a wide range of stylistic variations, thereby enriching the process of template extraction with a multitude of organizational perspectives. Secondly, despite this diversity, all requirements are composed in English and adhere to widely recognized grammatical and syntactic standards. Such uniformity facilitates the identification of recurring templates within different contextual frameworks.

Before the template induction, we manually reviewed the requirements and found that some were too complex or too simple to capture clear intentions. Firstly, we found that some requirements spanned multiple sentences. Since our goal in this study is to extract and recommend a single SRL sequence from one requirement, we focused solely on requirements expressed in single sentences, filtering out multi-sentence requirements first. Additionally, some requirements were too simple to convey clear semantics, being incomplete or too ambiguous. For instance, statements like ``Create test results'' are challenging to interpret accurately, raising questions such as what kind of test, who creates the results, and how to present them. These unclear requirements often have the characteristic of being very short. We manually filtered out these ambiguous requirements as well which typically appear as simple verb-object phrases. Two of our authors conducted paired annotation and reviewed each requirements. After applying these two types of filtering, 768 requirements remained for our analysis.

We then employed AllenNLP to process these 768 requirement statements, generating the semantic tags for each constituent unit such as word or phrase in each requirement.  Consider the requirement ``\emph{The system shall require a smart card reader, smart card, and a PIN to digitally sign an order}''. The SRL annotation is ``\emph{The system{\footnotesize[Arg0]} shall require{\footnotesize[REL]} a smart card reader{\footnotesize[Arg1]}, smart card{\footnotesize[Arg1]}, and a PIN{\footnotesize[Arg1]} to digitally sign{\footnotesize[ArgM-PRP]} an order{\footnotesize[ArgM-PRP]}}'', culminating in the SRL tag sequence \emph{[Arg0][REL][Arg1][Arg1][Arg1][ArgM-PRP][ArgM-PRP]}. Upon concluding this stage, we identified a total of 157 recurring sequences.



Two of our authors reviewed the 157 sequences and identified overlap and redundancy among them, culminating in three key categories of conclusions.
\begin{itemize}[leftmargin = 0.3cm]
   \item \textbf{Consistent structure with varied modal verbs}. Certain sequences exhibit a consistent structure, wherein the principal variation lies in the employment of modal verbs. Consider the sequences: ``[Arg0]must[V][Arg1]'', ``[Arg0]will[V][Arg1]'', and ``[Arg0]shall[V][Arg1]''.
    To reduce the number of final templates and increase the precision of template recommendations in later phases, we have standardized the use of modal verbs by adopting ``shall'' as the singular form, i.e., ``[Arg0]shall[V][Arg1]''. This choice aligns with various requirement standards that recommend using ``shall'' \cite{8559686, 5328509}. In practice, engineers can effortlessly modify the modal verb if differentiation is necessary.

    \item \textbf{Analogous syntactic structures with divergent semantic elements.} Certain sequences exhibit analogous syntactic structures, yet they contain elements that diverge semantically, not fitting neatly within a uniform semantic category. Take, for instance, the expressions ``[Arg0]should[V][Arg1][ArgM-TMP]'' and ``[Arg0]shall[V][Arg1][ArgM-PRP]'' Here, [ArgM-TMP] signifies temporal context, while [ArgM-PRP] indicates the purpose of an action—two distinct semantic dimensions.

    When faced with such variability, we simplify the template by preserving the shared structure and introducing a variable part. The resulting template, ``\emph{[Arg0]shall[V][Arg1][variable part]}'', accommodates either [ArgM-TMP] or [ArgM-PRP] in the [variable part].
    
    
    \item \textbf{Proper prefix.} A sequence may be a proper prefix of another, for example, ``\emph{[Arg1]must[V]}'' and ``\emph{[Arg1]must[V][Arg1]}''. In such cases, we streamline the pattern by retaining the common prefix and classifying the extended portion as a variable element. Consequently, the two sequences coalesce into a unified template ``\emph{[Arg1]must[V][variable part]}'', with the [variable part] representing the additional [Arg2]. Besides, we will standardize the modal verb ``must'' by consistently replacing it with ``shall'' to ensure uniformity across our generation.
   
\end{itemize}

    Finally, two variable SRL templates were derived from these 768 requirements, as shown in Table \ref{tab:template}. Each template contains two variable parts, denoted by `*'. These variable parts are optional, meaning that for each requirement sentence, each of these variable parts can be either present or absent. In other words, `*' indicates either zero or one occurrence.

    Additionally, we identified eight SRL tags that commonly appear across the 768 analyzed requirements, as delineated in Table \ref{tab:commonSRL}. These fine-grained tags can be used to fill in the variable parts of the template. Specifically, the [variable part] may include any arrangement of Arg2 and the ArgM types listed in Table \ref{tab:commonSRL}. More precisely, it can encompass any combination of Arg2, ArgM-PRP, ArgM-TMP, ArgM-LOC, and ArgM-BNF in various sequences. Consequently, under the assumption that each tag occurs only once, these templates can produce a minimum of 1,304 unique permutations. In reality, the same tags may appear multiple times, significantly increasing the number of permutations that these two flexible templates can accommodate. Thus, the generality of these templates is notable, despite being extracted from a dataset comprising only 768 requirements. The extent of their applicability and effectiveness will be thoroughly assessed in Section \ref{subsubsec:templateImpact}.

\begin{table}[!htbp]
\centering
\footnotesize
\caption{Two variable templates induced from 768 requirements.}
\renewcommand{\arraystretch}{1.5}
\label{tab:template}
\begin{tabular}{c|p{8cm}}
\hline
\textbf{ID} & \textbf{Template} \\ \hline
1 & [Arg2]*[Arg0][shall][V][Arg1][variable part]* \\ 
2 & [Arg2]*[Arg1][shall][V][variable part]* \\ \hline
\multicolumn{2}{p{13cm}}{Note: Each template contains two variable parts denoted by *. The * indicates that the variable part is optional, meaning it can be either present or absent.}
\end{tabular}
\end{table}


\begin{table}[!htbp]
\centering
\footnotesize
\caption{Eight high-frequency SRL tags.}
\label{tab:commonSRL}
\begin{tabular}{c|c|p{12.5cm}}
\hline
\textbf{ID} & \textbf{Name} & \textbf{Explanation}   \\ \hline
1 & REL(V) & Predicate \\ 
2 & Arg0 & Typically represents the ``Agent'' or ``Experiencer'' of an action.\\ 
3 & Arg1 &  Typically signifies the entity undergoing the action or change of state or the theme of the verb.   \\
4 & Arg2 &  Its meaning can vary widely depending on the verb but frequently represents the ``Beneficiary'', ``Instrument'', or ``End point'' roles.  \\ 
5& ArgM-PRP & Indicates the purpose or reason behind an action.\\
6& ArgM-TMP & Specifies the time when an action happens.\\
7 & ArgM-LOC &  Indicates the location where an action takes place. \\
8 & ArgM-BNF & Denotes an action that is done to benefit someone else.\\
\hline
\end{tabular}
\end{table}

\subsection{Phase II: Template-variant Recommendation}
\label{subsec:TRECTrain}

This phase aims to recommend a SRL template variant that effectively structures the sequence of semantic elements in a FR description to align with the specified feature. This involves ensuring all variables within the template are accurately instantiated with details from the feature. Thus, this process comprises two sub-tasks: first, selecting a variable SRL template, and second, mapping the tokens of the feature to the corresponding placeholders in the chosen variable template's structure. For instance, as shown in Fig.\ref{fig:MotiExample}, for the given feature ``RBC controls train movement,'' our recommended SRL template is ``[Arg0] RBC [Arg0] shall [V] control [V] [Arg1] train movements [Arg1].''

The challenge arises from the gap between the diverse and abstract lexical sequences in the features and their semantic roles in more concrete FR statements. Feature representations can differ, encompassing short sentences such as ``RBC controls train movement,'' verb-object phrases like ``sending videos'' \cite{9793551}, or even single or dual noun phrases, for instance, ``overlapped window layout'' \cite{kang_1990}. To bridge the gap between these varying representations and the precise semantic roles, we convert feature descriptions into pairs of (tokens, syntactic roles) in accordance with the syntactic role definitions provided by the FR syntax of ISO 29148.

This strategy rests on two pivotal observations. First, each feature can typically be articulated as a clear subject-predicate-object sentence. For verb-object constructs, we imply `the system' as the subject by default. For single noun phrases (e.g., overlapped window layout), we treat them as objects in statements such as ``the system should have an overlapped window layout.'' Given this simple construction of feature representations, current NLP tools, including Stanford CoreNLP \cite{manning-EtAl:2014:P14-5} and NLTK\footnote{\url{https://www.nltk.org/}}, can efficiently execute this standardization process. Secondly, our selection of the ISO 29148 FR syntax is deliberate; its conformity with common sentence structures allows the natural subject-verb-object flow to match the subject-action-object framework used in FR syntax.

\begin{figure*}[htbp]
	\centering
	\includegraphics[trim={0cm 4cm 0cm 0cm}, clip, width=\textwidth]{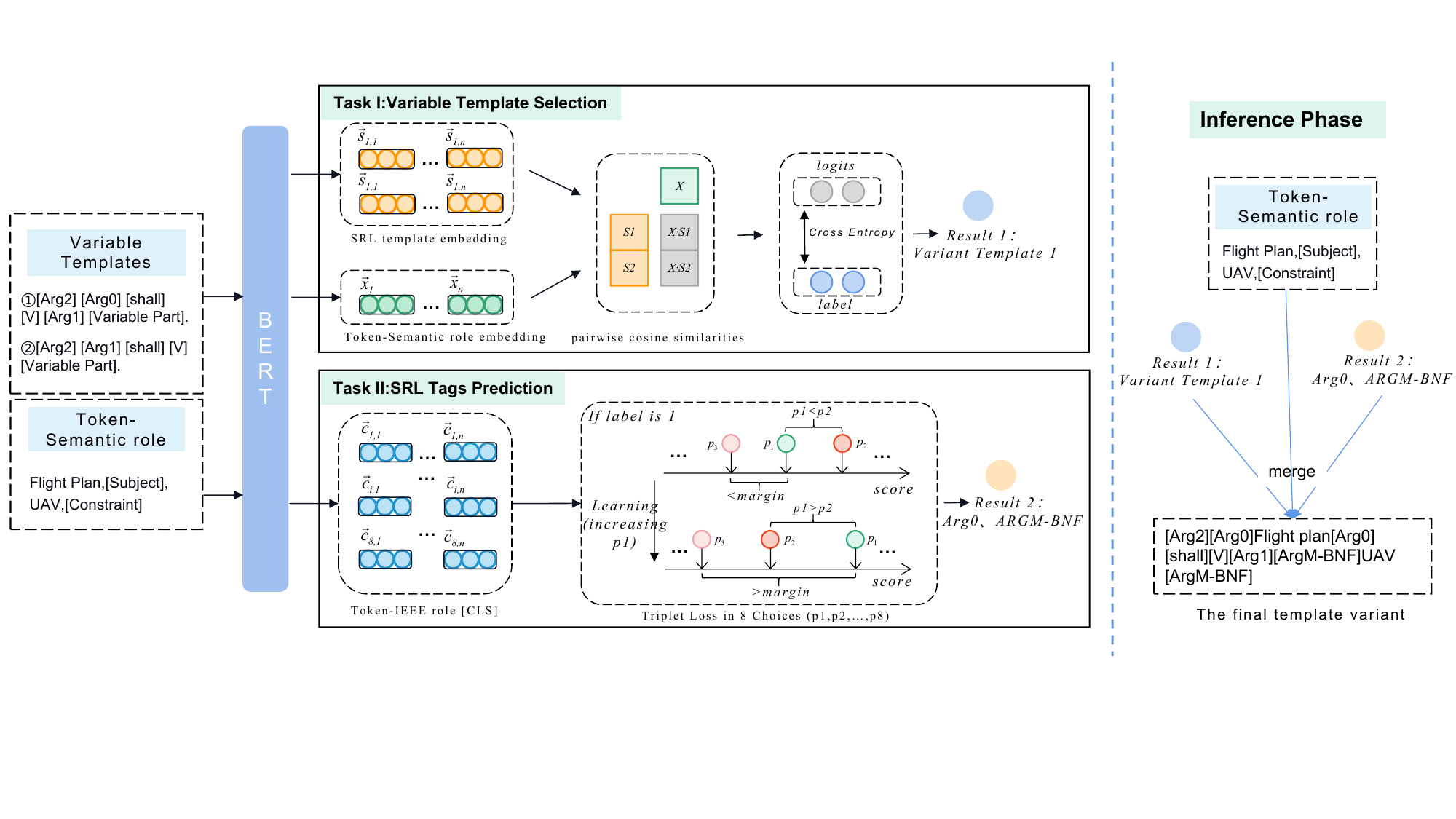}
	\caption{The training architecture of \textit{Key2Temp}.}
	\label{fig:STRM}
\end{figure*}

Consequently, the aim of this phase shifts to recommending an SRL template variant based on the pairs of tokens and syntactic roles, derived from one particular feature. To accomplish this, we have developed the \textit{Key2Temp} model. \textit{Key2Temp} is purpose-built as a dual-task model to tackle these sub-tasks, utilizing the BERT framework \cite{devlin-etal-2019-bert}. BERT is selected for its bidirectional attention mechanism and MLM, which enable it to accurately comprehend the semantic information of text. Additionally, BERT has demonstrated advantages in text classification tasks \cite{DBLP:conf/kbse/ZhaoY0LY024, DBLP:journals/spe/ZhaoZLL24}.
The model's architecture and function are depicted in fig. \ref{fig:STRM}.

\noindent \textbf{Task I: Variant Template Selection.} For the first task of variable template prediction, we borrow the idea of model CLIP (Contrastive Language-Image Pre-training) \cite{CLIP}, which was initially proposed to predict the correct pairings of the given batch of (image, text). It contrastively learns the features of images and text, and synthesizes a simple linear classifier to identify the text for a given image.

Similarly, in our task, the pairs of (keyword, syntactic role) and variable SRL templates represent different forms of a requirement description statement, just like the relations of image and its descriptive text for a given image. We regard these two as two modalities, and embed them with BERT independently firstly. Then, through a dropout layer and a fully connected layer, we obtain a probability matrix by calculating the cosine similarity between their encoding. For a list of input pairs of (keyword, syntactic role), the variable SRL template with the highest probability score is selected, which is the result of the first task. The cross entropy is selected as the loss function.

\noindent \textbf{Task II: SRL Tags Prediction.} The second task focuses on predicting SRL tags for specified tokens. As per the above discussion, we have eight potential choices (see Table \ref{tab:commonSRL}), turning this into a multi-class classification problem. We employ BERT to embed each pair of (keyword, syntactic role) and derive the CLS tokens. We then process these vectors with a dropout layer and a fully connected layer before predicting the tag with the output probability matrix. Triplet Loss is utilized as the loss function with the optimization objective of minimizing the distance between positive pairs (i.e., anchor and positive samples) while maximizing the distance between negative pairs (i.e., anchor and negative samples). As illustrated in the lower part of Fig. \ref{fig:STRM}, which represents the loss calculation process for a pair of (keyword, syntactic role). P1-P8 represent the predicted scores for 8 different common SRL tags in Table \ref{tab:commonSRL}. Taking the example in Fig. \ref{fig:STRM}, if the label is 1, then P1 is the positive tag, and P2-P7 are considered negative tags. The essence of the training process is to adjust the distance between the negative tags and P1 and make them larger than the predefined margin.

\noindent \textbf{SRL Template Variant Construction during Inference.} During inference, we embed SRL tags (results of task II) into the predicted variable template (result of task I), by locating the fixed tag and then inserting the variable tag, as the final template variant to guide the later requirements generation.  
For instance, for the input `` \emph{Flight plan[Subject], UAV[Constraint]}'', the predicted variable template is ``\emph{[Arg2][Arg0][shall][V][Arg1][Variable part]}''. The predicted SRL tags for the two tokens are \emph{Arg0} and \emph{ArgM-BNF}, respectively. Here, \emph{Arg0} is a fixed tag, and \emph{ArgM-BNF} falls under the variable part. We then insert \emph{Flight plan} after \emph{Arg0} and append \emph{UAV} at the end. To differentiate the content of one tag from others during parsing, we incorporate the tag name before and after each keyword. Consequently, the final template becomes \emph{[Arg2][Arg0]Flight plan[Arg0][shall][V][Arg1][ArgM-BNF]UAV[ArgM-BNF]}.

During implementation, we need to prepare the training dataset for \emph{Key2Temp}, which includes tokens, their syntactic roles, and the corresponding template variants. 
Due to the lack of a public available repository, we designed a reverse engineering process to distill the necessary tokens and syntactic roles from  public requirements specifications.
In the template induction phase (Section \ref{subsec:tempalteReduction}), we successfully derived SRL tags for all potential tokens and developed a semantic template for each of the 768 requirements. In other words, tokens and their template variants were identified. The following step involves participants annotating these tokens with ISO syntactic roles. Corresponding to the number of features, we randomly selected 2-5 tokens, such as verbs, verb phrases, nouns, or noun phrases, from each requirement.

We enlisted 12 computer science students and provided them with introductory training on the ISO 29148 syntax. We elucidated each syntactic role by presenting two illustrative examples and conducted discussions to ensure their understanding was solidified. It was agreed that identifying the ISO syntactic roles of the given tokens is intuitive, as they are almost consistent with the grammar of typical English sentences. Therefore, we did not perform the independent annotation-joint discussion procedure. Instead, the students were paired for annotation. Specifically, the annotation workflow proceeded as follows: (1) Requirements and extracted tokens were shared via Feishu\footnote{\url{https://docs.feishu.cn/docs}} (a collaborative editing platform); (2) Participant pairs conducted annotations either online through Tencent Meeting\footnote{\url{https://www.tencent.com/en-us/responsibility/combat-covid-19-tencent-meeting.html}} or in offline face-to-face sessions; (3) For any disagreements, a third-party arbitrator (typically one of our authors) mediated discussions until full consensus was reached on all annotation decisions.


\subsection{Phase III: Draft Generation}
\label{subsec:draftGeneration}

As illustrated in Fig. \ref{fig:overview}, we first fine-tuned ChatGLM-6B \cite{team2023chatglm} using a dataset that includes pairs of tokens and their ISO syntactic roles, the recommended SRL template variants, and the reference requirement statements. The purpose of this fine-tuning is to familiarize ChatGLM-6B with our task, specifically the relationships between pairs of tokens and syntactic roles, as well as the SRL template variants. During inference, we injected the template variant recommended by \emph{Key2Temp} into ChatGLM-6B with the aim of improving its ability to understand the semantic relationships and relative positions of the given tokens in the final requirement description, as well as enhancing the grammar of the generated sentences.
Our ChatGLM-6B fine-tuning employs P-Tuning V2 \cite{liu-etal-2022-p}, chosen for its memory efficiency (6GB with INT4 quantization), human-aligned outputs\footnote{\url{https://chatglm.cn/blog}}, and parameter-efficient adaptation (Section \ref{sec:background}).



We deliberately chose not to fine-tune ChatGLM-6B on the 768 manually annotated requirements from the \emph{Key2Temp} training set. Although these requirements contain correct SRL templates, they do not reflect the real-world inference scenario, where the model generates drafts based on automatically suggested SRL templates—which may contain errors. Fine-tuning on pristine manual annotations could introduce a distributional mismatch between training and inference data, potentially degrading model performance.

This approach aligns with findings from Zan et al. \cite{zan2023privatelibraryoriented}, who demonstrated that large language models can achieve better performance when fine-tuned on noisy, realistic data rather than immaculate annotations. Thus, we opted to use the recommended templates (with inherent noise) paired with token batches during fine-tuning, ensuring the model adapts to actual deployment conditions.


\section{Evaluation}
\label{sec:eva}

The goal of this section is to evaluate the extent to which our \emph{EasyFR} can generate feasible FRs from given features. To achieve this, we consider the two typical scenarios that requirements engineers may face when using assistance tools like \emph{EasyFR} to aid in the specification of software requirements.

\begin{itemize}[leftmargin=0.5cm]
    \item  \textbf{Scenario I}: \textit{EasyFR} is charged with generating entirely new requirements in cases where no existing related requirements are available.

    \item \textbf{Scenario II}: There's an existence of some requirements from the current or similar product, and \textit{EasyFR} is utilized to further generate requirements. 
\end{itemize}

    Therefore, we assess the performance of \textit{EasyFR} in these two contexts. We design three research questions (RQs) for this evaluation. The first two RQs are addressed under Scenario I: RQ1 provides an overview evaluation, and RQ2 is an ablation study to assess the impact of SRL template recommendations on FR generation. RQ3 focuses on the efficacy of our approach under Scenario II.

\begin{itemize}[leftmargin=0.5cm]
    \item \textbf{RQ1: How effective is \textit{EasyFR} at generating FRs in the absence of any pre-existing product requirements?} (Scenario I)

    \item  \textbf{RQ2: To what extent does the SRL template recommendation aid in the creation of functional requirements within \textit{EasyFR}?} (Scenario I)
    
    \item  \textbf{RQ3: How effectively does \textit{EasyFR} perform in generating FRs when fine-tuned with a subset of existing product requirements?} (Scenario II)
\end{itemize}


\subsection{Datasets}
\label{subsec:datasets}

The absence of publicly accessible feature repositories and their corresponding requirements poses a significant barrier to the evaluation of our approach. To navigate this constraint, we employ a suboptimal approach that draws from the wealth of openly available functional requirements.

The objective of this study is to generate FRs directly from features. Faced with the unavailability of ideal datasets, we have devised a method that emulates the typical process through which FRs are formulated based on their corresponding features.

We select representative tokens from existing FRs and use these as surrogates for the types of tokens commonly linked with features. This unconventional yet strategic methodology is a practical solution to the constraints imposed by our dataset limitations. Given that features are often encapsulated as minimal sentences, verb-object phrases, or noun phrases, we have deliberately confined both the number of tokens and their respective parts of speech.

To ensure a focused and realistic simulation, we limit our token selection to between 3 and 5 terms and assess three prevalent token combinations: 
\textbf{T1}: A combination comprising one token from the subject, one from the object, and 1 to 3 additional tokens from other parts of the requirement.
\textbf{T2}: A blend that includes one token from the subject, one from the predicate, and an additional 1 to 3 tokens from various segments.
\textbf{T3}: A selection of 3 to 5 tokens chosen at random.

The four open datasets employed for evaluation include UAV requirements \cite{8444851}, HIPAA requirements \cite{HIPPA}, ETCS requirements \cite{ETCS} and BAS \cite{BASReq}. Detailed information on these datasets can be found in Table \ref{tab:datasets}. Particularly, the UAV requirements stem from Dronology, an open-source Unmanned Aerial System (UAV) developed by the University of Notre Dame. These requirements, totaling 99, describe the functionalities for managing and coordinating the flight of small UAVs. HIPAA corresponds to a compilation of 1889 patient healthcare product requirements, collected from diverse sources including open-source products, IT healthcare standards, and commercial product feature descriptions. Detailed information and source references for these ten datasets can be found in work \cite{HIPPA}. The ETCS dataset, comprising 136 functional requirements, has been established to delineate the specifications for the European Train Control System (ETCS). Additionally, the City of Toronto uniformly implements a standard Building Automation System (BAS) specification across all new constructions, retrofits, and upgrades within its facilities, encompassing 460 functional requirements.

Please note that eight datasets from HIPAA were utilized to train the \textit{Key2Temp} model (refer to Section \ref{subsec:tempalteReduction}). During this evaluation phase, we selected token combinations that were new—those not used in the training of \textit{Key2Temp}.

\begin{table*}[!htbp]
	\centering
	\footnotesize
	\caption{The information on the four datasets used in this work.}
	\label{tab:datasets}
	\begin{tabular}{|c|p{12cm}|c|}
		\hline
        \textbf{Dataset}   &\textbf{Description} & \textbf{\#Req} \\ \hline
        UAV \cite{8444851}  &The requirements for an open-source Unmanned Aerial System outline the functionalities needed to manage and coordinate UAV flights. &99\\ \hline
        HIPAA \cite{HIPPA} & The Software Requirements Specifications (SRS) for ten patient healthcare products are included, with requirements gathered from diverse sources such as open-source products, IT healthcare standards, and feature descriptions of commercial products. &1889\\ \hline
        ETCS \cite{ETCS} & The requirements defines the FRs for EUROPEAN RAIL TRAFFIC MANAGEMENT SYSTEM / EUROPEAN TRAIN CONTROL SYSTEM. &136\\ \hline
        BAS \cite{BASReq}&Building Automation System (BAS) specification describes the central building automation system components and network protocol specifications. &460\\ \hline
        \multicolumn{2}{|c|}{Overall} &2584 \\ \hline
        \end{tabular}
\end{table*}

\subsection{Baselines}
\label{subsec:baseline}

\begin{itemize}[leftmargin = 0.5cm]
    \item \textbf{Bidirectional and Auto-Regressive Transformers (BART)} \cite{DBLP:journals/corr/abs-1910-13461}, developed by Facebook AI, is a sequence-to-sequence model pre-trained as a denoising autoencoder. It takes both left and right context into account, excelling in various text generation tasks. We selected it as a baseline because it has been evaluated as a robust approach for the software requirements generation task \cite{ReqGen}. We use the BART-large model with 400M parameters.

    \item  \textbf{GENIUS} \cite{guo2022genius} is an innovative conditional text generation model designed to use sketches as input and fill in missing contexts. A sketch comprises key textual elements such as phrases, spans, or words, which are concatenated using mask tokens. It is evident that GENIUS was designed to address tasks similar to ours. Furthermore, GENIUS displays superiority in terms of text generation quality, when compared to other notable conditional language models (CLMs). Therefore, it serves as another robust baseline for our research. We select the largest version with 406M parameters.
    
    \item  OpenAI's \textbf{Generative Pre-trained Transformer 4 (GPT-4-0613)} \cite{DBLP:conf/nips/BrownMRSKDNSSAA20} represents an enhancement over the initial release of GPT-4. Exhibiting superior text interpretation and generation capabilities, it has become an essential benchmark for generative tasks, including the current one at hand.
\end{itemize}

\begin{figure}[htbp]
	\centering
	\includegraphics[trim = {0cm 6.5cm 6cm 0cm}, clip, width=0.8\textwidth]{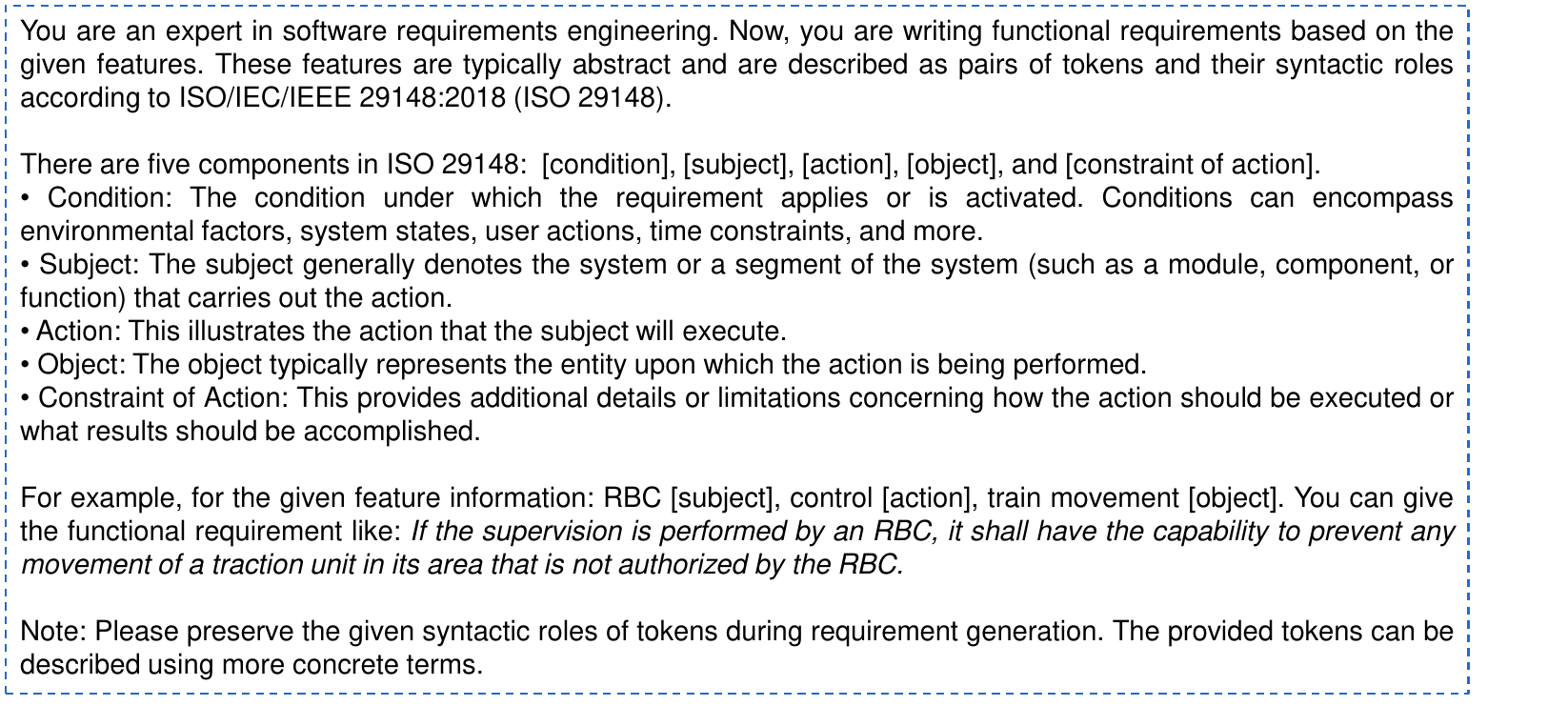}
	\caption{The prompt used for GPT-4-0613 in this study.}
	\label{fig:prompt}
\end{figure}

BART and GENIUS were fine-tuned using the same datasets as our \emph{EasyFR} during the following experiments. Additionally, GPT-4 was utilized with dialogue prompts. The GPT-4 prompt is shown in Fig. \ref{fig:prompt}.

\subsection{Metrics}
\label{subsec:metrics}

To evaluate the quality of the generated samples, we adopt the most commonly used Bilingual Evaluation Understudy (BLEU) \cite{papineni2002bleu}, as per the methodology presented in \cite{10172729}. BLEU scores are calculated by comparing n-grams within the generated sentences to the n-grams within the reference answers. Specifically, we utilize BLEU-2, BLUE-3, and BLEU-4 metrics. Additionally, we utilize METEOR \cite{lavie-agarwal-2007-meteor}, which significantly mitigates one of BLEU's notable shortcomings: its unreliability in producing consistent sentence-level evaluations. As a more robust metric, METEOR enhances assessment accuracy by incorporating syntax and stem matching as well as explicit word order alignment. Moreover, we adopt the NIST metric \cite{NIST}, a derivative of BLEU that leverages information entropy to weight matches. This approach de-emphasizes trivial phrases, thus diminishing the overrepresentation of high-frequency words in evaluation outcomes. Therefore, METEOR and NIST serve as more robust and significant metrics in our study.

\subsection{Addressing RQ1-RQ2 (Scenario I)}

To rigorously evaluate both the effectiveness and generalizability of our approach, we adopt a robust evaluation strategy: in Phase III, we fine-tune ChatGLM-6B on the UAV and HIPAA datasets while reserving ETCS and BAS for testing (applying the same procedure to BART). This 2+2 split (rather than 3+1) provides three key advantages: (1) it enables comprehensive assessment of cross-domain generalization through dual independent test sets; (2) it mitigates overfitting risks inherent in limited-domain fine-tuning of large language models; and (3) it better reflects real-world deployment scenarios where models must handle diverse, potentially imperfect inputs.

\vspace{0.5em} 
\subsubsection{Addressing RQ1: Efficacy of \emph{EasyFR}}
\label{subsubsec:RQ1}

The evaluation results are presented in Table \ref{tab:scenario1}, where we display the average values of each metric. We have structured the table according to token combinations and enhanced readability by formatting the best indicator values in bold and the second-best in italics for each metric across each token combination zone. 

To evaluate performance differences between each baseline and our proposed \emph{EasyFR}, we conduct one-tailed Mann-Whitney U tests, a distribution-free method requiring no assumptions about data distribution. The alternative hypothesis ($H_{1}$) states that \emph{EasyFR}'s performance is statistically superior to the baseline under a one-tailed test. Results showing statistically significant superiority of our \emph{EasyFR} ($p$ < 0.05) are annotated with **, while marginally significant results (0.05 $\leq p <$ 0.1) are marked with *. And the strong and weak significant superiority of the baseline are annotated with $^{- -}$ and $^{-}$. We apply Holm-Bonferroni correction for multiple comparisons. Complete test results including adjusted $p$-values, Cohen's $d$ effect sizes, and statistical power appear in Table \ref{tab:significance-Scenario1}.

\begin{table*}[!htbp]
	\centering
	\footnotesize
	\caption{Models' results in scenario I (\%). }
	\label{tab:scenario1}
	\begin{tabular}{|c|c|ccccc|ccccc|}
		\hline
		\multirow{2}{*}{\textbf{Tokens}} 	& \multirow{2}{*}{\textbf{Approach}} & \multicolumn{5}{c|}{\textbf{ETCS}} & \multicolumn{5}{c|}{\textbf{BAS}} \\ 
        \cmidrule(lr){3-7} \cmidrule(lr){8-12}
	    &&\textbf{METEOR} & \textbf{NIST} & \textbf{BLEU-2} & \textbf{BLEU-3} & \textbf{BLEU-4} & \textbf{METEOR} & \textbf{NIST} & \textbf{BLEU-2} & \textbf{BLEU-3} & \textbf{BLEU-4}  \\ \hline
	\multirow{4}{*}{T1}  
    & GENIUS &33.92** &109.23** &37.13** &31.79** &27.76**       &30.46** &92.84** &35.58** &29.26** &24.76**  \\
	& BART &40.26 &182.15 & 53.12 &49.47 &46.48    &33.28** &117.64** &41.83** &38.83** &35.62**  \\ 
    & GPT-4 &\textbf{46.73} &\textbf{226.89} & \textbf{60.96} &\textbf{56.56} &\textbf{53.00}    &\emph{34.53}** &\emph{139.33}** & \emph{48.45} &\textbf{44.36} & \textbf{41.28}  \\ 
	&EasyFR &\emph{44.10} &\emph{210.87} & \emph{58.90} & \emph{53.75} & \emph{49.83}  &\textbf{36.48} &\textbf{162.65} & \textbf{49.38} &\emph{43.99} &\emph{40.02} \\ \hline \hline

		\multirow{4}{*}{T2} 
        & GENIUS &33.16** &103.25** & 38.12** &31.89** &27.37**     &29.27** &83.05** &34.85** &27.86** &23.05** \\ 
		& BART &34.69** &161.83** & 45.91** &42.37* & 39.67     &28.85** &100.26** &36.20** &32.56** &29.88**  \\ 
        & GPT-4 &\emph{40.28} &\emph{165.03}** & \emph{46.59}** & \emph{44.33}** & \emph{42.51}     &\emph{31.30}** &\emph{122.63} &\emph{44.07} &\textbf{39.88} &\textbf{36.92} \\ 
		&EasyFR &\textbf{40.93} &\textbf{197.22} &\textbf{54.33} & \textbf{48.65} & \textbf{44.63} &\textbf{33.90} &\textbf{143.45} &\textbf{44.99} &\emph{39.67} &\emph{35.91}   \\ \hline \hline
		
		\multirow{4}{*}{T3} 
        & GENIUS &34.34** &111.42** &37.96** &32.16** &27.85**     &29.60** &85.37**  &34.68** &28.00** &23.37**  \\ 
		& BART &39.52 &172.28* & 50.30* &45.61 &41.95      &30.59** &101.87** &38.05** &34.26** &31.46** \\ 
        & GPT-4 &\textbf{46.58} &\textbf{208.62} & \textbf{57.59} &\textbf{54.16$^{-}$} &\textbf{51.41$^{- -}$} 
        &\emph{33.61}* &\emph{145.49} &\textbf{48.11} & \textbf{43.28$^{-}$} & \textbf{39.81$^{- -}$}   \\ 
		&EasyFR &\emph{43.36} &\emph{202.91} & \emph{56.52} & \emph{50.52} & \emph{46.13} &\textbf{35.40} &\textbf{150.37} & \emph{46.91} &\emph{41.60} &\emph{37.79} \\ \hline 
        \multicolumn{12}{l}{**: Strong significance of our EasyFR ($p<0.05$); *: Weak significance of our EasyFR ($0.05 \leq p<0.1$)}\\
        \multicolumn{12}{l}{$^{- -}$: Strong significance of GPT-4 ($p<0.05$); $^{-}$: Weak significance of our GPT-4 ($0.05 \leq p<0.1$)}
	\end{tabular}
\end{table*}

\begin{table*}[!htbp]
  \centering
  \footnotesize
  \caption{Significance test results (adjusted $p$-value ($p\_Holm$), Cohen’s $d$, and power) for EasyFR vs. Baselines.}
  \label{tab:significance-Scenario1}
  \setlength{\tabcolsep}{1.8pt}
  \begin{tabular}{|c|c|c|ccc|ccc|ccc|ccc|ccc|}
    \hline
    \multirow{2}{*}{\textbf{Dataset}}&\multirow{2}{*}{\textbf{Tokens}} & \multirow{2}{*}{\textbf{Baseline}} 
      & \multicolumn{3}{c|}{\textbf{METEOR}} 
      & \multicolumn{3}{c|}{\textbf{NIST}} 
      & \multicolumn{3}{c|}{\textbf{BLEU-2}} 
      & \multicolumn{3}{c|}{\textbf{BLEU-3}} 
      & \multicolumn{3}{c|}{\textbf{BLEU-4}} \\
    \cline{4-18}
    &  & & $p\_Holm$ & $d$ & Power 
        & $p\_Holm$ & $d$ & Power 
        & $p\_Holm$ & $d$ & Power 
        & $p\_Holm$ & $d$ & Power 
        & $p\_Holm$ & $d$ & Power \\
    \hline
    \multirow{9}{*}{ETCS} &\multirow{3}{*}{T1}
     & GENIUS 
        & 0 &  0.75& \cellcolor{blue!20}1.00 
        & 0 &  1.14& \cellcolor{blue!20}1.00 
        & 0 &  1.35 & \cellcolor{blue!20}1.00 
        & 0 &  1.43 & \cellcolor{blue!20}1.00 
        & 0 &  1.46 &\cellcolor{blue!20} 1.00 \\
     & & BART 
        & 0.14 & 0.27 & 0.73 
        & 0.17 & 0.26 & 0.68 
        & 0.11 & 0.29 & 0.78 
        & 0.26 & 0.23 & 0.59 
        & 0.35 & 0.20 & 0.50 \\
     & & GPT-4 
        & 1.00 & 0.10 & 0.20 
        & 1.00  & 0.17 & 0.39 
        & 1.00  & 0.20 & 0.49 
        & 1.00  & 0.24 & 0.62 
        & 1.00  & 0.25 & 0.67 \\
    \cline{2-18}
    
    &\multirow{3}{*}{T2}
      & GENIUS 
        & $8.0e{-5}$ &  0.55 & \cellcolor{blue!20}1.00 
        & 0 &  1.03 & \cellcolor{blue!20}1.00 
        & 0 &  1.05 & \cellcolor{blue!20}1.00 
        & 0 &  1.13 & \cellcolor{blue!20}1.00 
        & 0 &  1.19 & \cellcolor{blue!20}1.00 \\
    &  & BART 
        & $8.58e{-3}$ & 0.39 & \cellcolor{blue!20}0.95 
        & $3.89e{-2}$ & 0.32 & \cellcolor{blue!20}0.84 
        & $1.32e{-2}$ & 0.37 & \cellcolor{blue!20}0.92 
        & $5.05e{-2}$ & 0.29 & 0.78 
        & $0.11$ & 0.25 & 0.65 \\
    &  & GPT-4 
        & 0.35 & 0.10 & 0.21 
        & $3.89e{-2}$ & 0.32 & \cellcolor{blue!20}0.84 
        & $5.31e{-3}$ & 0.42 & \cellcolor{blue!20}0.96 
        & $1.06e{-2}$ & 0.24 & 0.63 
        & 0.25 & 0.12 & 0.26 \\
    \cline{2-18}
    
     &   \multirow{3}{*}{T3}
     & GENIUS 
        & 0  &  0.66 & \cellcolor{blue!20}1.00 
        & 0 &  1.15 & \cellcolor{blue!20}1.00 
        & 0 &  1.18 & \cellcolor{blue!20}1.00 
        & 0 &  1.25 & \cellcolor{blue!20}1.00 
        & 0 &  1.29 & \cellcolor{blue!20}1.00 \\
    &  & BART 
        & 0.24 & 0.24 & 0.63 
        & $9.73e{-2}$ & 0.30 & 0.79 
        & $9.90e{-2}$ & 0.29 & 0.77 
        & 0.24 & 0.23 & 0.61 
        & 0.34 & 0.20 & 0.50 \\
    &  & GPT-4 
        & 1.00 & 0.07 & 0.15 
        & 1.00 & 0.13 & 0.28 
        & 1.00 & 0.14 & 0.31 
        & 1.00 & 0.28 & 0.74 
        & 1.00 & 0.36 & \cellcolor{blue!20}0.91 \\
 \hline
 \hline

  \multirow{9}{*}{BAS}&\multirow{3}{*}{T1}
      & GENIUS 
        & 0 & 0.49 & \cellcolor{blue!20}1.00 
        & 0 &  0.85 & \cellcolor{blue!20}1.00 
        & 0 &  0.90 & \cellcolor{blue!20}1.00 
        & 0 &  1.03 & \cellcolor{blue!20}1.00 
        & 0 &  1.11 & \cellcolor{blue!20}1.00 \\
     & & BART 
        & $7.00e{-5}$ & 0.33 & \cellcolor{blue!20}1.00 
        & 0 &  0.50 & \cellcolor{blue!20}1.00 
        & 0 & 0.42 & \cellcolor{blue!20}1.00 
        & $3.00e{-5}$ & 0.34 & \cellcolor{blue!20}1.00 
        & $4.30e{-4}$ & 0.29 & \cellcolor{blue!20}0.99 \\
    &  & GPT-4 
        & $2.69e{-3}$  & 0.25 & \cellcolor{blue!20}0.96 
        & $2.69e{-3}$  & 0.25 & \cellcolor{blue!20}0.96 
        & 0.95 & 0.04 & 0.13 
        & 1.00 & 0.02 & 0.09 
        & 1.00 & 0.07 & 0.24 \\
    \cline{2-18}
     &   \multirow{3}{*}{T2}
      & GENIUS 
        & 0  & 0.43 & \cellcolor{blue!20}1.00 
        & 0  &  0.76 & \cellcolor{blue!20}1.00 
        & 0  &  0.67 & \cellcolor{blue!20}1.00 
        & 0  &  0.83 & \cellcolor{blue!20}1.00 
        & 0  &  0.94 & \cellcolor{blue!20}1.00 \\
    &  & BART 
        & 0  & 0.47 & \cellcolor{blue!20}1.00 
        & 0  & 0.49 & \cellcolor{blue!20}1.00 
        & 0  & 0.46 & \cellcolor{blue!20}1.00 
        & 0  & 0.41 & \cellcolor{blue!20}1.00 
        & 0   & 0.38 & \cellcolor{blue!20}1.00 \\
     & & GPT-4 
        & $2.00e{-5}$  & 0.31 & \cellcolor{blue!20}1.00 
        & 0.86  & 0.26 & \cellcolor{blue!20}0.99 
        & 1.00 & 0.04 & 0.15 
        & 1.00 & 0.02 & 0.10 
        & 1.00 & 0.07 & 0.27 \\
      \cline{2-18}
      &   \multirow{3}{*}{T3}
      & GENIUS 
        & 0  & 0.48 & \cellcolor{blue!20}1.00 
        & 0  &  0.80 & \cellcolor{blue!20}1.00 
        & 0  &  0.79 & \cellcolor{blue!20}1.00 
        & 0  &  0.93 & \cellcolor{blue!20}1.00 
        & 0  &  1.03 & \cellcolor{blue!20}1.00 \\
     & & BART 
        & 0  & 0.42 & \cellcolor{blue!20}1.00 
        & 0 &  0.54 & \cellcolor{blue!20}1.00 
        & 0  & 0.43 & \cellcolor{blue!20}1.00 
        & 0   & 0.40 & \cellcolor{blue!20}1.00 
        & 0   & 0.38 & \cellcolor{blue!20}1.00 \\
     & & GPT-4 
        & 0.08  & 0.15 & 0.71 
        & 0.97& 0.05 & 0.18 
        & 1.00 & 0.11 & 0.51 
        & 1.00 & 0.14 & 0.71 
        & 1.00 & 0.17 & \cellcolor{blue!20}0.82 \\
     \hline
     \multicolumn{18}{l}{Cohen's $d$ thresholds: 0.2 (small), 0.5 (medium), 0.8 (large). Power $\geq$0.8 is recommended.}\\ 
     \multicolumn{18}{l}{Color coding indicates effect size and statistical power: blue background = strong statistical power.} \\ 
  \end{tabular}
\end{table*}

We can have the following observations from Table \ref{tab:scenario1} and \ref{tab:significance-Scenario1}.
\begin{itemize}[leftmargin=0.5cm]
    \item \textbf{Both GPT-4 and our \emph{EasyFR} demonstrate strong performance, with \emph{EasyFR} achieving superior results in the primary metrics of METEOR and NIST.} 
    As evidenced in Table \ref{tab:scenario1}, our proposed \emph{EasyFR} consistently outperforms both GENIUS and BART across all experimental conditions. Statistical significance testing (Table \ref{tab:significance-Scenario1}) confirms that \emph{EasyFR} demonstrates statistically significant superiority over GENIUS ($p <$ 0.05) for all token combinations and evaluation benchmarks.

    In comparisons with GPT-4, \emph{EasyFR}  shows competitive advantages in specific configurations. For the ETCS dataset, it achieves better performance with the T2 token combination (comprising one subject token, one predicate token, and one token from any other segment). In the BAS dataset, \emph{EasyFR}  outperforms GPT-4 in the two primary metrics of METEOR and NIST across all token combinations. However, GPT-4 yields stronger results in BLEU-3 and BLEU-4. The significance testing indicates that GPT-4's advantages are limited to the T3 token combination (random tokens), where it shows weak significance in BLEU-3 and strong significance in BLEU-4 across both datasets. This finding actually has limited practical relevance, as real-world applications typically involve scenarios where engineers have approximate knowledge of subject-predicate relationships, corresponding to the T2 combination, rather than dealing with completely random token arrangements.
    In contrast, \emph{EasyFR}  demonstrates strong significance in METEOR all token combinations in the BAS dataset. Notably, all significant results for \emph{EasyFR}  exhibit statistical power exceeding 0.80, underscoring the robustness of these findings.

    This performance pattern suggests that while GPT-4 excels at precisely matching 3-grams and 4-grams in reference answers, \emph{EasyFR} generates more flexible variations through strategies such as synonym substitution, morphological word variations, and alternative grammatical sequences. Given that neither system consistently produces exact matches to reference FRs, \emph{EasyFR}'s flexibility proves more practical for real-world applications.

    Remarkably, \emph{EasyFR} achieves this competitive performance despite its substantially smaller scale, 6 billion parameters compared to GPT-4's estimated 1.8 trillion\footnote{\url{https://www.semianalysis.com/p/gpt-4-architecture-infrastructure}}. This efficiency is particularly evident in its strong performance on METEOR and NIST, even as it shows marginally lower scores in BLEU-3 and BLEU-4.

    \item \textbf{GPT-4 tends to integrate all provided tokens into fluent English sentences without rigorously adhering to the specified syntactical roles, despite clear instructions in the prompt.}  For instance, faced with the input ``actuator, 0, operators, 1, manually position, 2'' (where the numbers 0-2 indicate syntactic roles according to ISO 29148, as explained in Section \ref{subsec:functionalReqSyntax}: 0 indicates a condition, 1 indicates a subject, and 2 indicates an action),  GPT-4 constructs the sentence ``\emph{Operators shall be able to manually position actuators in the event of system failure},'' neglecting our guidelines on the syntax hint of \emph{actuator} which results in omitting the precondition from the requirement. This tendency explains why GPT-4 often achieves higher BLEU scores but falls short in METEOR and NIST assessments. In contrast, \emph{EasyFR} treats syntactic constraints with strict adherence, constructing coherent sentences within these confines. This indicates that although GPT-4 achieves higher BLEU-3/BLEU-4, its generated requirements drafts may need more modifications, since it would miss blocks of syntax units.

    \item \textbf{The smaller models, GENIUS and BART, lag behind in performance.} Nevertheless, it is worth noting BART's commendable output, achieved with a mere 400M parameters, demonstrating the potential of smaller models when larger ones are not feasible. 
    However, occasionally, \emph{BART tends to produce responses that are excessively literal, echoing the input tokens too closely and resulting in text that feels unnatural.} For example, given the input ``It, 1, a train stop, 3,'' (where the numbers 1 and 3 indicate the syntactic roles of subject and object, respectively), BART might simply return ``It, 1, a train stop.'' This problem of verbatim responses is observed in 13.24\% and 9.93\% of cases within two datasets for T2 keyword combinations, respectively.

    This behavior aligns with BART's pre-training objective to reconstruct original text from corrupted versions, hence its predilection for retaining all inputs during reconstruction. This suggests that further enhancements are necessary for BART, especially in scenarios with limited computational resources.

    \item \textbf{All models demonstrated enhanced performance with ETCS specifications while showing reduced effectiveness on BAS.} Through comparative analysis of these two datasets, the inherent limitations of the automated models on FR generation become apparent.     Firstly, the BAS dataset is replete with acronyms which obscure the interpretation of tokens due to the density of specialized terms. Secondly, BAS requirements often feature compound sentences that encapsulate multiple directives or coordinated clauses, posing a significant challenge for models to predict more intricate semantics. This bottleneck might be alleviated by enriching the input with more guiding tokens. Finally, the generation of BAS-specific requirements frequently necessitates domain-specific expertise. 
    
    To illustrate these points, consider the requirement: ``\emph{Each AAC shall reside on a BACnet network using ISO 8802-3 (Ethernet) Data Link/Physical layer protocol with BACnet/IP addressing, or it shall reside on a BACnet network using the MS/TP Data Link/Physical layer protocol.}'' Crafting such a specification necessitates a detailed comprehension of the network protocols relevant to AAC, ISO 8802-3 Data Link/Physical layer protocol, and the intricacies of the BACnet infrastructure, thus highlighting the complexity that current models encounter when dealing with content that is rich in domain-specificity and articulated through complex sentence structures.
\end{itemize}

\vspace{0.1cm}
\begin{mdframed}
[linecolor = gray!100,linewidth = 3pt,
innerleftmargin = 3pt, topline=false, rightline=false, bottomline=false, leftline=true, innerrightmargin = 3pt,innertopmargin = 3pt, innerbottommargin = 3pt,backgroundcolor = gray!30]
\textbf{Summary for RQ1:} 
Our study reveals that GPT-4 and our \emph{EasyFR} outperform the other two baselines on both datasets. Moreover, our \emph{EasyFR} excels in the two significant metrics: METEOR and NIST. Although GPT-4 shows an advantage in BLEU-3 and BLEU-4, it tends to integrate all tokens provided into fluent English sentences while ignoring the given syntactical roles restrictions. This means that more effort is needed to modify the generated drafts.
\end{mdframed}

\vspace{0.5em} 
\subsubsection{Addressing RQ2: Impact of SRL template recommendation.}
\vspace{0.5em} 
\label{subsubsec:templateImpact}

To evaluate the impact of SRL template guidance on token-driven requirement generation, we conducted an ablation study comprising three key experiments: (1) fine-tuning EasyFR with only original tokens and syntactic roles (no template), (2) augmenting the input with the ISO 29148 template (simplified as `ISO template'), and (3) integrating a recommended SRL template variant. This systematic removal and reintroduction of template components isolates their individual contributions to model performance. Here, the ISO template refers to the sequence of ISO 29148 syntactic roles occurring in a specific requirement statement, such as \emph{[subject][action][object]} and so on.
The ISO template was selected for its authoritative guidance on structuring FRs. 

Our evaluation of \textit{EasyFR} spans nine scenarios, combining three token combinations (T1, T2, and T3) with three different template settings, i.e., no template, the ISO template, and our suggested SRL template variant. Table \ref{tab:templateImpact} displays these outcomes. For clarity, the best metric values are in bold, and variations from one row to the next are indicated by arrows: upward ($\uparrow$) for improvements and downward ($\downarrow$) for declines, exemplified by comparing the performance of the ISO template + T1 against T1 alone, without any template usage. 

\begin{table*}[!htbp]
	\centering
	\footnotesize
	\caption{Impact of SRL template recommendation for \emph{EasyFR} in scenario I.}
	\label{tab:templateImpact}
	\begin{tabular}{|c|ccccc|ccccc|}
		\hline
		\multirow{2}{*}{\textbf{tokens/template}} & \multicolumn{5}{c|}{\textbf{ETCS}} & \multicolumn{5}{c|}{\textbf{BAS}} \\ 
        \cmidrule(lr){2-6} \cmidrule(lr){7-11} 
		& \textbf{METEOR} & \textbf{NIST} & \textbf{BLEU-2} & \textbf{BLEU-3} & \textbf{BLEU-4}  
        & \textbf{METEOR} & \textbf{NIST} & \textbf{BLEU-2} & \textbf{BLEU-3} & \textbf{BLEU-4}   \\ \hline 
		
		T1 & 42.62 &209.39 &57.89 &52.06 &47.57  &32.82 &122.52 &43.86 &39.57 &36.32 \\ 
		ISO Template+T1 & 42.31\tiny$\downarrow$ &199.48\tiny$\downarrow$ &57.43\tiny$\downarrow$ &51.92\tiny$\downarrow$ &47.73\tiny$\uparrow$  
        &32.80\tiny$\downarrow$ &120.69\tiny$\downarrow$ &43.31\tiny$\downarrow$ &38.92\tiny$\downarrow$ &35.58\tiny$\downarrow$ \\
		SRL Template+T1 &\textbf{44.10}\tiny$\uparrow$&\textbf{210.87}\tiny$\uparrow$ &\textbf{58.90}\tiny$\uparrow$& \textbf{53.75}\tiny$\uparrow$& \textbf{49.83}\tiny$\uparrow$  
        &\textbf{36.48}\tiny$\uparrow$ &\textbf{162.65}\tiny$\uparrow$ & \textbf{49.38}\tiny$\uparrow$ &\textbf{43.99}\tiny$\uparrow$ &\textbf{40.02}\tiny$\uparrow$ \\ \hline \hline
		
		T2 &40.16 &191.52 &54.97 & 48.47 & 43.84    &30.91** &118.37 &41.28 &36.14 &32.44   \\ 
		ISO Template+T2 & \textbf{42.05}\tiny$\uparrow$ &196.00\tiny$\uparrow$ & 54.13\tiny$\downarrow$ &48.15\tiny$\downarrow$ &43.85\tiny$\uparrow$  
        &31.00\tiny$\uparrow$ &117.94\tiny$\downarrow$ &40.70\tiny$\downarrow$ &35.85\tiny$\downarrow$ &32.33\tiny$\downarrow$ \\ 
		SRL Template+T2 &40.93\tiny$\uparrow$&\textbf{197.22}\tiny$\uparrow$ & \textbf{54.33}\tiny$\downarrow$ & \textbf{48.65}\tiny$\uparrow$ & \textbf{44.63}\tiny$\uparrow$  
        &\textbf{33.90}\tiny$\uparrow$ &\textbf{143.45}\tiny$\uparrow$ &\textbf{44.99}\tiny$\uparrow$ &\textbf{39.67}\tiny$\uparrow$ &\textbf{35.91}\tiny$\uparrow$  \\ \hline \hline
			
		T3 & 43.06 &196.02 &55.60&49.87&45.28 &31.44 &115.57 &41.59 &36.92 &33.46   \\ 
		ISO Template+T3 &\textbf{45.66}\tiny$\uparrow$ &\textbf{204.93}\tiny\tiny$\uparrow$ & 56.04\tiny$\uparrow$ &50.31$\uparrow$ & 45.93\tiny$\uparrow$ 
        &31.83\tiny$\uparrow$ &117.77\tiny$\uparrow$  &40.45\tiny$\downarrow$ &36.02\tiny$\downarrow$ &32.75\tiny$\downarrow$ \\ 
		SRL Template+T3 &\emph{43.36}\tiny$\uparrow$&\emph{202.91}\tiny$\uparrow$ &\textbf{56.52}\tiny$\uparrow$ & \textbf{50.52}\tiny$\uparrow$& \textbf{46.13}\tiny$\uparrow$ 
        &\textbf{35.40}\tiny$\uparrow$ &\textbf{150.37}\tiny$\uparrow$ & \textbf{46.91}\tiny$\uparrow$ &\textbf{41.60}\tiny$\uparrow$ &\textbf{37.79}\tiny$\uparrow$\\ \hline 
	\end{tabular}
\end{table*}

Upon examining the table, it becomes clear that \textit{EasyFR}, when utilizing our recommended SRL template variants, achieves the best metric scores for all token groupings in 27 out of 30 metrics across all token configurations. Furthermore, \textit{EasyFR} equipped with the SRL template invariably surpasses the version without any template.  This indicates:

\begin{itemize}[leftmargin = 0.4cm]
    \item \textbf{Our SRL templates refine the precision of generated requirements.} The adoption of our SRL templates effectively transforms a open and complex NLG task into a more structured slot-filling exercise. The uniform improvement across all metrics indicates that this approach enables ChatGLM-6B to predict tokens more accurately within designated slots. 
    
    \item \textbf{The robustness and generalizability of SRL template recommendations.} Although the SRL templates are derived from a mere 768 requirements in only three domains: healthcare, project management and web archive (refer to Section \ref{subsec:tempalteReduction})—their application extends successfully to other areas, such as ETCS and BAS.

    \item \textbf{The ISO templates positively impact METEOR and NIST scores for T2 and T3 tokens across both datasets.} However, its application slightly hampers the performance for T1 tokens in these datasets. This observation suggests that expansive syntactic roles defined by the ISO template, where one ISO role may comprise multiple SRL tags, may not provide significant guidance for generating FRs. Instead, they might introduce noise that disrupts the original template-free generation process.
\end{itemize}

\vspace{0.1cm}
\begin{mdframed}
[linecolor = gray!100,linewidth = 3pt,
innerleftmargin = 3pt, topline=false, rightline=false, bottomline=false, leftline=true, innerrightmargin = 3pt,innertopmargin = 3pt, innerbottommargin = 3pt,backgroundcolor = gray!30]
\textbf{Summary for RQ2:} 
The study reveals that the use of SRL templates can enhance all metrics across both datasets. In contrast, ISO templates only improve the METEOR and NIST metrics for token combinations T2 and T3, indicating that more fine-grained semantic roles are needed to guide FR sentence generation.
\end{mdframed}

\subsection{Addressing RQ3: Effectiveness of \textit{EasyFR} in Scenario II}
\label{subsubsec:scenario2}

In scenario II, we aim to assess the effectiveness of \textit{EasyFR} if it has been fine-tuned with a set of pre-existing requirements from a particular product. Owing to the notable performance of BART and GPT-4 in Scenario I (as depicted in Table \ref{tab:scenario1}), we elected to use them as our baselines in Scenario II. 

To increase the stability of the results, we employ n-fold cross-validation, where n is usually set to 5 or 10. Theoretically, the number of folds k is approximately set to $\log(n)$ , where n is the number of samples. In our case, n is 2584, and $\log(n)$ is about 3.41. Thus, we select the typical 5 folds.
Additionally, computational resources are a significant consideration when setting k. More folds mean more experiments and greater computational demands of fine-tuning. Even with 5-fold cross-validation, each fold requires 4-5 hours on a GPU 2080 on average with our datasets, resulting in the entire cross-validation experiment taking over 100 hours.
In summary, we select 5-fold cross-validation for our experiment.

For each of the four datasets, i.e., UAV, HIPAA, ETCS and BAS, the requirements are partitioned into five equal segments. Four of these segments are utilized to fine-tune ChatGLM-6B and BART, while the remaining segment serves as the testing portion. During these procedures, we ensure that each segment has been used as the testing part once. In evaluating the algorithm, we compute average values across all metrics on each dataset. Additionally, we calculate the gains our \textit{EasyFR} has made in each metric, with $\uparrow$ signifying positive growth and $\downarrow$ denoting decrease. The outcomes are presented in Table \ref{tab:scenario2}. 

Consistent with the analysis in Table \ref{tab:scenario1}, we employ one-tailed Mann-Whitney U tests to assess statistical significance. Results are annotated as follows: ** and * indicate statistically strong ($p <$ 0.05) and weak ($0.05 \leq p <$ 0.10) superiority of \emph{EasyFR} over baseline methods, respectively, while $^{--}$ and $^{-}$ denote corresponding levels of baseline superiority over \emph{EasyFR}.

Main observations include:
\begin{itemize}[leftmargin = 0.2cm]

\item \textbf{Pre-existing requirements bolster the performance of our \textit{EasyFR} in generating requirements.} By comparing with the results in scenario I, results indicated in Table \ref{tab:scenario2} clearly show that \textit{EasyFR} outshines its previous performance in Scenario I in both datasets. Furthermore, it exhibits more significant gains in the ETCS dataset, with improvements of 13.28\%, 17.51\%, 20.95\%, 20.66\%, and 30.67\% across various metrics. However,  the advances in the BAS are less pronounced due to the inherent complexity of its requirements, as discussed in Section \ref{subsubsec:RQ1}. A similar trend is noticeable in BART's performance.

\item Different from the weaker performance of \textit{EasyFR} than GPT-4 in Scenario I in the metrics of BLEU-3 and BLEU-4, \textbf{our \textit{EasyFR} surpasses GPT-4 in every metric across all datasets, except for UAV when very limited pre-existing requirements can be used for training.} Notably, our approach achieves strong significantly better results than GPT-4 in 13 out of the 15 comparisons across these three benchmarks, except the BLEU-3 and BLEU-4 in BAS. The greatest margins of improvement over GPT-4 are observed in HIPAA, boasting the most extensive training set with 1,280 requirements, followed by the ETCS dataset (108 training requirements).

\item \textbf{GPT-4 significantly surpasses \textit{EasyFR} in performance on the UAV dataset} due to two main factors. First, the scant number of pre-existing training requirements available for \textit{EasyFR}, specifically only 80 for the UAV context, limits its potential for improvement. Second, unlike the other three datasets, this one is compiled by academics and precisely structured according to the Easy Requirements Specification \cite{5328509}. As a result, the requirements are exceptionally clean and straightforward. In such scenarios, the advanced capabilities of GPT-4 enable it to predict sentence structures accurately, even without leveraging our detailed SRL-based templating approach.

\item \textbf{Both BART and our \textit{EasyFR} yield the strongest results on the ETCS dataset.} Despite the HIPAA dataset being the largest used for fine-tuning (comprising 1280 requirements), the superior metrics are achieved with ETCS. This suggests that, beyond the sheer volume of the fine-tuning dataset, the syntax consistency of individual requirements also play significant roles. The HIPAA dataset amalgamates ten datasets from diverse sources, including open-source hospital information systems, healthcare standards, and requirement exemplars, among others. Consequently, variations are notable in terminology usage, requirement syntax, and representation. In contrast, ETCS is the product of a single organization, and its requirements have undergone several rounds of review and modification. Therefore, comparatively speaking, learning from and re-utilizing the knowledge embedded in pre-existing ETCS requirements should be easier.

Furthermore, with the limited training data in ETCS (108 requirements), we observe that EasyFR's performance advantage shows no statistical significance compared to Scenario I. This aligns with consistently low statistical power values (all below 0.8) across evaluation metrics, indicating insufficient testing cases to reliably detect significant differences. A similar pattern emerges in the UAV dataset, where small sample sizes likewise limit statistical conclusiveness.
\end{itemize}

\begin{threeparttable}[!htbp]
\centering
\footnotesize
\caption{Models' results in scenario II. }
\label{tab:scenario2}
\begin{tabular}{|c|c|ccccc|}
\hline
\textbf{Data.} & \textbf{Approach} & \textbf{METEOR} & \textbf{NIST}  & \textbf{BLEU2}  &\textbf{BLEU3} &\textbf{BLEU4}    \\ \hline

\multirow{2}{*}{UAV} 
& BART &45.28 \tiny $\downarrow$1.30\% &228.74 \tiny$\downarrow$4.49\% 
& 60.16 \tiny $\downarrow$1.86\%\tnote{1} &56.54 \tiny$\downarrow$0.76\% &53.70 \tiny$\uparrow$0.02\% \\ 

&GPT-4 &\textbf{50.62$^{-}$} \tiny $\uparrow$9.38\% &\textbf{262.72$^{-}$} \tiny $\uparrow$9.02\%  
& \textbf{65.60$^{- -}$} \tiny $\uparrow$6.59\%\tnote{2} &\textbf{61.66$^{- -}$} \tiny$\uparrow$7.61\% & \textbf{58.57$^{- -}$} \tiny $\uparrow$8.33\% \\ 

&\emph{EasyFR} &45.87 &239.02 &61.28 &56.97 &53.69  \\ \hline \hline

\multirow{2}{*}{HIPAA} 
& BART & 40.06** \tiny $\downarrow$20.19\% & 182.60** \tiny $\downarrow$31.53\% 
& 52.17** \tiny $\downarrow$18.40\% & 49.15** \tiny $\downarrow$17.92\% & 46.69** \tiny $\downarrow$18.06\%  \\ 

&GPT-4 &36.00** \tiny $\downarrow$33.75\% & 155.90** \tiny $\downarrow$53.91\% 
& 49.53** \tiny $\downarrow$24.71\% & 44.05** \tiny $\downarrow$31.58\% & 40.12** \tiny $\downarrow$37.39\% \\ 

&\emph{EasyFR} &\textbf{48.15} &\textbf{240.18} &\textbf{61.77} &\textbf{57.96} &\textbf{55.12} \\ \hline \hline

\multirow{2}{*}{ETCS} 
& BART &49.56 \tiny $\downarrow$7.36\% &257.49 \tiny $\downarrow$7.01\% 
& 62.93 \tiny $\downarrow$6.02\% &59.77 \tiny $\downarrow$5.67\% &57.18 \tiny $\downarrow$5.40\% \\ 

&GPT-4 &46.73** \tiny $\downarrow$13.87\% &226.89** \tiny $\downarrow$21.44\% 
&60.96** \tiny $\downarrow$9.45\% &56.56** \tiny $\downarrow$11.67\% &53.00** \tiny $\downarrow$13.72\% \\ 

&\emph{EasyFR} &\textbf{53.21} & \textbf{275.54}  &\textbf{66.72}&\textbf{63.16} & \textbf{60.27}  \\ \hline \hline

\multirow{2}{*}{BAS} 
& BART &33.07** \tiny $\downarrow$12.76\% &122.26** \tiny $\downarrow$37.44\%  
& 41.22** \tiny $\downarrow$24.31\% &37.63** \tiny $\downarrow$22.03\% &34.88** \tiny $\downarrow$20.41\% \\  

&GPT-4 &34.53** \tiny $\downarrow$7.99\% &139.33** \tiny $\downarrow$20.61\% 
& 48.45** \tiny $\downarrow$5.76\% &44.36 \tiny $\downarrow$3.52\% &41.28 \tiny $\downarrow$1.74\% \\ 

&\emph{EasyFR} &\textbf{37.29} &\textbf{168.04} &\textbf{51.24} &\textbf{45.92} &\textbf{42.00}     \\ \hline

\end{tabular}
\begin{tablenotes}
    \tiny
    \item[1]This reflects the BLEU-2 gain achieved by \textit{EasyFR} over UAV. $\downarrow$ indicates that the approach in the corresponding row performs worse, while \textit{EasyFR} works better.
    \item[2]This reflects the BLEU-2 gain achieved by \textit{EasyFR} over GPT-4. $\uparrow$ indicates that the approach in the corresponding row performs better, while \textit{EasyFR} works worse.
    \item[3] Symbols ** and * denote statistically strong and weak significant superiority (p < 0.05 and 0.05$\leq$p$<$0.1) of \emph{EasyFR} over the baseline, respectively. Conversely, -- represents strong significant superiority and - represents weak significant superiority of baseline methods over EasyFR.
\end{tablenotes}
\end{threeparttable}

\begin{table*}[!htbp]
  \centering
  \footnotesize
  \caption{Significance test results (adjusted $p$-value($p\_Holm$), Cohen’s $d$, and power) for \emph{EasyFR} vs. Baselines.}
  \label{tab:significance-uav-hipaa}
  \setlength{\tabcolsep}{1.8pt}
  \begin{tabular}{|c|c|ccc|ccc|ccc|ccc|ccc|}
    \hline
    \multirow{2}{*}{\textbf{Data.}} & \multirow{2}{*}{\textbf{Baseline}} 
      & \multicolumn{3}{c|}{METEOR} 
      & \multicolumn{3}{c|}{NIST} 
      & \multicolumn{3}{c|}{BLEU-2} 
      & \multicolumn{3}{c|}{BLEU-3} 
      & \multicolumn{3}{c|}{BLEU-4} \\
    \cline{3-17}
      & & $p\_Holm$ & $d$ & Power 
        & $p\_Holm$ & $d$ & Power 
        & $p\_Holm$ & $d$ & Power 
        & $p\_Holm$ & $d$ & Power 
        & $p\_Holm$ & $d$ & Power \\
    \hline
    \multirow{2}{*}{UAV}
      & BART 
        & 1.00 & 0.16 & 0.30 
        & 1.00 & 0.04 & 0.08 
        & 1.00 & 0.01 & 0.06 
        & 1.00 & 0.03 & 0.08 
        & 1.00 & 0.04 & 0.09 \\
      & GPT-4 
        & 1.00 & 0.22 & 0.47 
        & 1.00 & 0.26 & 0.57 
        & 1.00 & 0.36 & \cellcolor{blue!20}0.81 
        & 1.00 & 0.36 & \cellcolor{blue!20}0.81 
        & 1.00 & 0.36 & \cellcolor{blue!20}0.82 \\
    \hline
     \hline
    \multirow{2}{*}{HIPAA}
      & BART 
        & 0.00  & 0.34 & \cellcolor{blue!20}1.00 
        & 0.00   & 0.45 & \cellcolor{blue!20}1.00 
        & 0.00   &  0.53 & \cellcolor{blue!20}1.00 
        & 0.00   & 0.49 & \cellcolor{blue!20}1.00 
        & 0.00   & 0.46 & \cellcolor{blue!20}1.00 \\
      & GPT-4 
        & 0.00   & 0.59 & \cellcolor{blue!20}1.00 
        & 0.00   &  0.68 & \cellcolor{blue!20}1.00 
        & 0.00   &  0.59 & \cellcolor{blue!20}1.00 
        & 0.00   &  0.65 & \cellcolor{blue!20}1.00 
        & 0.00   &  0.68 & \cellcolor{blue!20}1.00 \\
    \hline \hline
    
    \multirow{2}{*}{ETCS}
      & BART 
        & 0.33 & 0.19 & 0.47 
        & 0.38 & 0.13 & 0.28 
        & 0.33 & 0.19 & 0.47 
        & 0.38 & 0.15 & 0.33 
        & 0.38 & 0.12 & 0.25 \\
      & GPT-4 
        & 0.04  & 0.31 & \cellcolor{blue!20}0.81 
        & $1.72e{-3}$  & 0.45 & \cellcolor{blue!20}0.98 
        & $2.34e{-3}$  & 0.34 & \cellcolor{blue!20}0.88 
        & $1.39e{-3}$  & 0.37 & \cellcolor{blue!20}0.92 
        & $1.26e{-3}$  & 0.38 & \cellcolor{blue!20}0.93 \\
    \hline
    \hline
    \multirow{2}{*}{BAS}
      & BART 
        & $4.00e{-5}$  & 0.33 & \cellcolor{blue!20}1.00 
        & 0.00  & 0.49 & \cellcolor{blue!20}1.00 
        & 0.00  &  0.52 & \cellcolor{blue!20}1.00 
        & 0.00  & 0.47 & \cellcolor{blue!20}1.00 
        & 0.00  & 0.44 & \cellcolor{blue!20}1.00 \\
      & GPT-4 
        & $8.40e{-4}$  & 0.27 & \cellcolor{blue!20}0.98 
        & $2.50e{-3}$  & 0.30 & \cellcolor{blue!20}0.99 
        & $4.48e{-2}$  & 0.15 & 0.66 
        & 0.19 & 0.10 & 0.39 
        & 0.19 & 0.07 & 0.23 \\
     \hline    
     \multicolumn{17}{l}{Cohen's $d$ thresholds: 0.2 (small), 0.5 (medium), 0.8 (large). Power $\geq$0.8 is recommended.} \\
     \multicolumn{17}{l}{Blue background = strong statistical power.}
  \end{tabular}
\end{table*}

\vspace{0.1cm}
\begin{mdframed}
[linecolor = gray!100,linewidth = 3pt,
innerleftmargin = 3pt, topline=false, rightline=false, bottomline=false, leftline=true, innerrightmargin = 3pt,innertopmargin = 3pt, innerbottommargin = 3pt,backgroundcolor = gray!30]
\textbf{Summary for RQ3:} 
It is evident that fine-tuning with pre-existing requirements can enhance the performance of our \emph{EasyFR}. The more fine-tuned datasets, the better the performance. In our study, with more than 100 requirements for fine-tuning across three datasets, our \emph{EasyFR} outperforms GPT-4 in all metrics.
\end{mdframed}

\section{Discussion}
\label{sec:discussion}

\subsection{Threats to validity}
\label{sec:threats}

\emph{Internal Validity}: While manual annotation of tokens with ISO 29148 syntactic roles may introduce errors, we mitigated this risk through paired annotation and discrepancy resolution procedure. Given the standard’s widespread adoption and alignment with English grammar conventions, we posit that any residual bias is likely minimal.

Furthermore, the SRL annotations from AllenNLP may present errors. However, our primary goal was to identify and recommend \emph{practically valid} SRL templates, those that are recognizable and actionable by contemporary models, not necessarily templates that are \emph{flawless}. The efficacy of these recommendations is substantiated by our experimental results, which signal their beneficial impact.

Due to the limitations of the annotated requirements (768 requirements) and the manual induction process, it is difficult to guarantee that we have obtained all possible templates. Additionally, it is challenging to ensure that our two variable templates can cover all requirements. However, the combinations of variable templates and configurable fine-grained tags significantly enhance the flexibility of our approach, providing a minimum of 1,304 permutations. Furthermore, the three experiments demonstrate the efficacy of our approach, with the SRL template serving as an essential component.

\noindent \emph{External Validity}: The external validity of our \textit{EasyFR} may first be affected by the evaluation dataset we used. In the absence of a publicly accessible repository for features, we simulated FR specifications from features and then reverse-engineered this process to glean feature token information from publicly available requirement statements. Although these tokens may not match the original features exactly due to their varied representations, they do, to a great extent, encapsulate the core tokens typical of features. This is due to two reasons: the advanced capabilities of current NLP tools allow for automatic transformation of features into tokens due to their straightforward expression. Additionally, we have observed the characteristic number of tokens and their parts of speech in feature cases presented in academic literature \cite{9793551}\cite{kang_1990}, which informed us in setting constraints on them.

The tokens and syntactic roles used in our experiments originated from established requirements, which might not correspond exactly with those in practice. Although we only tested three token combinations, \textit{EasyFR}'s consistent performance (comparable to GPT-4) suggests potential generalizability to practical scenarios, mitigating concerns about external validity.

While this work focuses on FRs, our approach is theoretically applicable to non-functional requirements (NFRs) (e.g., performance, security), as all requirement types follow natural language grammar and syntactic conventions. For example, performance requirements typically involve quantitative constraints on response time, throughput, and resource utilization. Future work can explore related performance.

Besides, the effectiveness of SRL template recommendation examined in this study has been limited to its implementation with ChatGLM-6B. Given the swift progress in AI models, future work should explore the applicability of SRL templates in requirements generation using a broader spectrum of language models.

\noindent \emph{Construct Validity}: To ensure valid measurement of FR generation quality, we utilize established NLG evaluation metrics (BLEU-2/3/4, METEOR, and NIST) that have been widely adopted in generative requirements engineering research \cite{10172729, math11020332}. Our assessment comprehensively covers both greenfield (from-scratch) and reference-based generation scenarios. Furthermore, we validate our approach through rigorous comparisons with diverse baselines, including specialized NLG models and general-purpose large language models. These carefully designed methodological decisions collectively establish strong construct validity for our approach's evaluation.

\noindent \emph{Conclusion Validity}: To ensure robust conclusion validity, we employed two key strategies. First, we conducted experiments on a sufficiently large sample (N=2,584) across four datasets, each constructed by independent research groups and spanning diverse domains. Second, to account for the non-normal distribution of our data and the limited sample sizes in the UAV and ETCS datasets, we employed the Mann-Whitney U test for significance testing and applied Holm-Bonferroni correction to control for multiple comparisons. The robustness of these statistical findings is further supported by our power analysis, which shows that four comparisons approached the conventional 0.8 threshold while all remaining comparisons exceeded this threshold across both experimental scenarios. 

\subsection{Limitations}
\label{subsec:limitations}

\begin{itemize}[leftmargin = 0.2cm]
    \item \emph{The refinement of overly abstract features is not involved.} Our focus is on generating a FR from its corresponding feature. Intuitively, the more information contained in the feature, the more feasible the generated FR will be. In our experiment, we set the token count between 3 to 5. According to Section \ref{subsubsec:RQ1}, given the token number limitation, effective tokens should at least include the subject and object, as verbs in natural language sentences are usually short and can be inferred from the subject and object. For overly abstract features, such as single verbs like ``copy,'' ``cut,'' and ``paste'' \cite{featureMei}, it is difficult to generate useful statements.

    \item  \emph{Single sentence generation limitation.}  Since our SRL templates are derived from individual sentences, our current approach is limited to producing a single sentence for each requirement based on one particular feature. The generation of requirements that need multiple sentences for further description is beyond the scope of our study.
    
    \item  \emph{ChatGLM-6B is not the cutting-edge backbone model.} Our choice of ChatGLM-6B was driven by computational resource limitations and its satisfactory performance. Nevertheless, its smaller scale may constrain overall performance. However, the principles and framework of SRL guiding FR generation are general. Future work can explore their application with more advanced models.
    
    \item  \emph{The practical implications remain unassessed}. While the experiments demonstrate that \emph{EasyFR} outperforms the baseline models, its actual impact on the efficiency of SRS and the quality of the resulting FRs has not been evaluated. Theoretically, due to the consistent framework of our approach across different requirements, \emph{EasyFR} should help eliminate the terminology inconsistency problem \cite{10.1145/3092703.3098239}. Additionally, by inducing complete semantic roles in high-quality SRS and recommending configured SRL template variants for given features, the generated requirements should mitigate the common issue of semantic incompleteness in single FRs \cite{DBLP:journals/ese/AroraSB19}. Future work can evaluate its practical significance in real-world scenarios.
    
\end{itemize}

\subsection{Implications}
\label{subsec:implications}

Our proposed approach, \emph{EasyFR}, and its findings hold significant implications for both academic research and industrial practitioners.

\subsubsection{Impact for Research.} This study establishes a novel research direction in generative software requirements engineering. While intelligent code synthesis has seen remarkable progress in recent years, automated requirements generation remains significantly underdeveloped. This gap presents a critical research challenge, as high-quality requirements serve as fundamental inputs for code synthesis. Current manual authoring processes are not only time-consuming but also prone to errors including terminology inconsistencies, incompleteness, and quality variability. As intelligent generative software engineering becomes an indispensable component of efficient development pipelines, the limitations of current approaches become increasingly apparent. While large language models (LLMs) can produce generic requirements, they consistently fail to generate outputs that properly adhere to project-specific contexts and required syntactic patterns. This fundamental constraint highlights the critical importance of developing effective mechanisms for injecting domain-appropriate constraints to enable high-quality requirements generation.

Our work pioneers semantic template-constrained generation for software requirements, but its implications extend far beyond this domain. The methodology offers a transferable framework for automating and standardizing critical technical documents, such as software design specification, product manual and patent application, where natural language precision and stylistic consistency are paramount. Our framework is transferable since these documents within most organizations usually follow similar same format, and the sentences present similar structure and pattern.

The demonstrated positive correlation between SRL template recommendations and the quality of generated natural language statements provides both theoretical grounding and empirical validation for this class of document generation systems.

\subsubsection{Impact for Practice.}

The strong performance of \emph{EasyFR}, particularly when fine-tuned on an organization’s historical requirements, demonstrates its potential for integration into domain-specific requirement IDEs. Although advanced coding models like GPT-4o \cite{GPT4}, Claude-3.5-sonnet \cite{Claude}, DeepSeek-Coder \cite{guo2024deepseek}, and Qwen-Coder \cite{qwenwebsite} and IDE like Cursor \footnote{\url{https://www.cursor.com/cn}} and Devin \footnote{\url{https://devin.ai/}} merges, the requirement models and IDEs are extremely rare. Considering that most organizations operate in focused domains and accumulate repositories of requirements with consistent language styles, it is possible to create an organization-individual requirements IDE by integrating the techniques like our \emph{EasyFR}.

\section{Related Work}
\label{sec:relatedWork}

We initially delve into research on requirements generation. Given that our task is fundamentally a constrained NLG task, demanding the inclusion of certain tokens in the final output, it is pertinent to also discuss relevant studies on constrained text generation.

\vspace{-1mm} 

Recent advancements in deep learning have spurred innovative applications in automated requirement generation. Arora et al. \cite{arora2023llms} systematically reviewed the potential of LLMs across different RE stages, highlighting their role in uncovering ``unknown requirements'', reducing ambiguities and facilitating human-machine interactions. Koscinski et al. \cite{10172729} explored the use of the Relational Generative Adversarial Network (RelGAN) to produce security requirements from functional requirements, highlighting RelGAN's practical utility through a case study. Zhao et al. \cite{ReqGen} generated requirement drafts by integrating domain ontology with the Unified pre-trained Language Model (UniLM), although their method may be hindered by the often-absent domain ontologies. Gudaparthi et al.\cite{10260978} concentrate on creativity within RE and suggest the generation of creative requirements using adversarial examples. This sets our research agendas apart. Our objective is to assist in articulating already understood requirements specifications for engineers, whereas their work proposes heuristic candidates, i.e., concepts that are not predetermined.


In the field of model-driven requirements engineering, various structured modeling approaches have been proposed to automate and refine the generation of textual specifications. For example, Cox et al. integrated business process models with Jackson’s Problem Frames to contextualize requirements in complex e-business projects \cite{turetken2004automating}, while Türetken et al. harnessed KAOS to automatically derive FRs from these process models, boosting efficiency and consistency \cite{DBLP:journals/infsof/CoxPBV05}. In parallel, Robertson et al. demonstrated the efficacy of i* models by applying 19 reusable patterns in an air traffic control setting, producing over 200 requirements statements \cite{DBLP:journals/re/MaidenMJG05}. Yu et al. introduced a dual-level framework that couples i* for organizational analysis with ALBERT \cite{Lan2019ALBERTAL} for system requirements, showcasing iterative refinement in a banking case study \cite{DBLP:conf/coopis/YuBDM95}. 

On the goal-oriented side, Letier and Lamsweerde combined KAOS and SCR for incremental modeling and precise behavior verification, employing a stepwise transformation to reconcile semantic and syntactic differences \cite{DBLP:conf/sigsoft/LetierL02}. Landtsheer et al. advanced this idea by defining formally proved operationalization patterns to map real-time temporal logic to software operations \cite{DBLP:journals/re/LandtsheerLL04}, while van Lamsweerde highlighted both the benefits and practical hurdles of transferring Goal-oriented requirements engineering (GORE) to industry \cite{van2004goal}. 

Further research on UML-based methods includes Berenbach’s technique for generating textual requirements and task lists from use case models \cite{DBLP:conf/re/Berenbach03a}, Meziane et al.’s GeNLangUML for producing natural language specifications \cite{DBLP:journals/re/MezianeAA08}, and RM2Doc for automatically generating ISO/IEC/IEEE 29148-compliant documents from UML models \cite{9793770}. Additionally, Lamsweerde and Willemet presented a learning-based approach to infer declarative KAOS goals from operational scenarios, enabling formal analyses such as conflict detection and obstacle analysis \cite{DBLP:journals/tse/LamsweerdeW98}. Despite their varying focuses, all these methods rely on precise modeling and carefully defined transformation rules, often necessitating domain expertise and ongoing model refinement. As the field evolves, further research is needed to enhance the flexibility, scalability, and domain adaptability of these methods, ensuring they remain practical and effective in dynamic real-world scenarios.

Constrained Text Generation (CTG) is a branch of NLG that aligns closely with our task, where text must adhere to predefined constraints. Notably, GENIUS leverages input sketches to flesh out missing context within a sketch \cite{guo2022genius}, demonstrating its utility for data augmentation in the automation of software requirement generation. For this study, we have chosen GENIUS as a benchmark model for comparison.
\section{Conclusion}
\label{sec:conclusion}

We present \textit{EasyFR}, a novel approach that utilizes configurable SRL template to enhance the generation of specific FRs from abstract features. In contrast to existing methods that hinge on domain ontologies \cite{ReqGen}, pre-existing collections of FRs \cite{10172729}, or (semi-)structured models \cite{turetken2004automating, DBLP:journals/infsof/CoxPBV05,DBLP:journals/re/MaidenMJG05, DBLP:conf/coopis/YuBDM95}, \emph{EasyFR} offers a more lightweight solution. By inducing recurring SRL templates from ten different requirements datasets and training our model of \emph{Key2Temp}, \emph{EasyFR} is able to suggest the most appropriate template variant for each distinct feature. The process includes selecting a variable template, slotting the given feature's tokens into their respective places, and structuring parts that need to be generated into a number of semantic slots, simplifying the typically complex task of generating requirements into a more manageable slot-filling activity. This results in well-aligned requirements with the provided features, while also maintaining consistency with the established patterns of FR expressions.
Our experiments demonstrate the effectiveness of \emph{EasyFR} both in scenarios where previous requirements exist and where they do not, highlighting the beneficial impact of our template recommendations.

Looking ahead, we intend to further refine the process of delineating overly abstract features (which can be divided into finer-grained features; for example, the feature \emph{edit} actually contains at least three features: \emph{copy}, \emph{cut}, and \emph{paste}) into FRs. We also aim to evaluate the practical value of our \emph{EasyFR} in real-world practice to enhance efficiency and mitigate potential issues such as semantic incompleteness and terminology inconsistency.


\section{Data Availability}
\label{sec:dataAvailability}

The implementation of \textit{EasyFR} and the associated dataset are available at \url{https://figshare.com/s/8fe6ad12f637b0a8923b}.

 \section*{Acknowledgments}

 Funding for this work was provided by the National Natural Science Foundation of China (Grants No. 62102014 and 62177003) and the National Science and Technology Major Project of China (Grant No. Y2022-V-0001-0027).
\bibliographystyle{ACM-Reference-Format}
\bibliography{Reference}

\end{document}